%% file: main.tex
\documentclass[12pt]{article}

\usepackage[T1]{fontenc}
\usepackage{kpfonts}
\usepackage{eurosym}
\usepackage{graphicx}
\usepackage{enumerate}
\usepackage{amsmath}
\usepackage{amsfonts}
\usepackage{amssymb}
\usepackage{amsfonts}
\usepackage{amssymb,amsmath,amsthm,epsfig,graphicx,natbib,subfigure,hyperref}
\usepackage{geometry}
\usepackage{caption}
\usepackage{color}
\usepackage{enumerate}
\usepackage{setspace}
\usepackage[textwidth=30mm]{todonotes}
\usepackage{sgame, tikz}
\usepackage{multirow,array}
\usepackage{rotating}
\usepackage{graphicx}
\usepackage{subfigure}
\usepackage{float}
\usepackage{booktabs}
\usepackage{makecell}
\usepackage{xurl}
\usepackage{threeparttable}
\usepackage{subcaption}
\usepackage{hyperref}
\usepackage{pgfplots}
\usepackage{soul}

\hypersetup{colorlinks=true,linkcolor=blue,citecolor=blue,urlcolor=blue}
\allowdisplaybreaks

\bibpunct[: ]{(}{)}{;}{a}{}{,}
\newtheorem{obs}{Observation}
\newtheorem{property}{Property}

\newtheorem{proposition}{Proposition}

\begin{document}

\input{article_2608}

\newpage

\part*{Online Appendix}

\appendix

\input{appendix}

\end{document}

%% file: article_2608.tex
\title{Artificial Intelligence, Data and Competition\thanks{This paper was previously circulated as ``Algorithmic Collusion and Price Discrimination: the Over-Useage of Data.'' We thank Giacomo Calzolari, Jean-Edouard Colliard, Francesco Decarolis, Bo Hu, Bruno Jullien, Sanxi Li, Tristan Tomala, Daniel Yi Xu, Jidong Zhou for their valuable comments and suggestions. We also thank seminar audiences at Fudan University, Nanjing University, Shandong University, University of International Business and Economics, the 2024 Beijing BEAT, the 2024 AMES, the Youth Forum of Digital Economics 2024, the Ninth PKU-NUS Annual International Conference on Quantitative Finance and Economics, the 2025 APIOC, the 2026 CUHK summer school. All errors are our own.}}

\author{Zhang Xu\thanks{School of Economics, Renmin University of China, China. Email:
xuzhang@ruc.edu.cn } \and Mingsheng Zhang\thanks{School of Economics, Renmin University of China, China. Email:
mingsheng\_zhang@yeah.net } \and Wei Zhao\thanks{
School of Economics and Management, Tsinghua University, China. Email:
wei.zhao@outlook.fr }}
\date{This Version: August 8, 2026\\[5pt] Latest Version: \href{https://365.kdocs.cn/l/cutf0Vnu3hYf}{Click Here}}

\maketitle

\begin{abstract}

This paper examines how data inputs shape competition among artificial intelligences (AIs) in pricing games. The dataset assigns labels to consumers and divides them into different market segments, thereby inducing multimarket contact among AIs. We document that AIs can adapt to tacit collusion via market allocation. Under symmetric segmentation, each algorithm monopolizes a subset of market segments with supra-competitive prices while competing intensely in the remaining market segments. Market segments with higher WTP are more likely to be assigned for collusion. Under asymmetric segmentation, the algorithm with finer segmentation adopts a Bait-and-Restraint-Exploit strategy to ``teach''  the other algorithm to collude. However, the data advantage does not necessarily result in competitive advantage. Our analysis calls for a close monitoring of the data selection phase, as the worst-case outcome for consumers can emerge even without any coordination. \\

\noindent \emph{Keywords:} {Consumer Labeling, Algorithmic Collusion, Price Discrimination, Multimarket Contact, Market Allocation}

\end{abstract}

\newpage
{
  \hypersetup{linkcolor=black}
  \tableofcontents
}

\clearpage
\pagenumbering{arabic}
\setcounter{page}{1}

\newpage

\section{Introduction}

A growing number of pricing decisions are delegated to artificial intelligence (AI). This trend is mainly driven by two properties of AI. On the one hand, AI can process massive volumes of consumer data and then conduct personalized pricing, regarded as price discrimination in narrow sense. On the other hand, AI can respond to variations in market conditions and rivals' strategies at high frequencies, regarded as price discrimination in broader sense.\footnote{See \cite{Bertini2021} and \cite{spann2024algorithmic}.} At the same time, concerns on both price discrimination and tacit collusion by AIs have been raised.\footnote{See \cite{oecd2017algorithmsandcollusion, oecd2018personalised}.} Both issues have been studied extensively but separately in the literature.\footnote{One of the most seminal works on price discrimination is \cite{bergemann2015limits} and on algorithmic collusion is \cite{Calvano2020}. A recently emerging literature on relation between price discrimination and competition, however, restricts attention to rational decision makers in static game (See \citealt{rhodes2024aer} and \citealt{Elliott2026}).} However, these two issues are not independent. Drawn based on studying separately, the prediction may not be robust or the policy implication may not hold, when these two issues coexist.\footnote{For instance, many policies on data, privacy and price discrimination may ignore its impact on competition among AIs. We will discuss in detail in Section \ref{sec:implication}.} In this paper, by considering these two issues at the same time, we study (1) how the consumers’ input dataset shapes competition between AIs and (2) how the competition affects personalized prices.

The fact that AI is adopted to offer personalized pricing by competing firms is prevalent in real life. In the ride-hailing sector, platforms such as Uber and Lyft leverage vast amounts of consumer data to implement AI-driven personalized pricing strategies. These algorithms allow firms to compete by dynamically personalizing prices.\footnote{See \url{https://www.nelp.org/ubers-price-gouging-and-what-we-can-do-about-it/}.} Competing consumer-lending platforms use AI to offer applicant-specific interest rates. Upstart uses AI and application data to determine loan rates,\footnote{See \url{https://www.upstart.com/learn/negotiate-a-personal-loan-rate/}.} while LendingClub uses machine learning to provide qualified applicants with customized rates.\footnote{See \url{https://www.sec.gov/Archives/edgar/data/1409970/000140997026000018/lc-20251231.htm}.} Furthermore, the Federal Trade Commission (FTC) has reported that at least 250 sellers employ surveillance pricing to charge different prices for the same goods based on a consumer’s location or browsing history.\footnote{While there is no explicit evidence that firms use surveillance pricing to compete directly, the large number of adopters implies that such competitive interaction is likely. See \url{https://www.ftc.gov/news-events/news/press-releases/2025/01/ftc-surveillance-pricing-study-indicates-wide-range-personal-data-used-set-individualized-consumer}.}

To formalize the role of consumer datasets in market competition, we model a dataset as a labeling system that assigns each consumer to a labeled segment. The labels may encode either payoff-relevant or payoff-irrelevant information. Because each firm's segmentation scheme is determined by its dataset, competing firms may segment consumers in the same or different ways. In each period, one consumer arrives, activating the segment associated with that consumer's label, and the two firms engage in Bertrand competition. Both firms use Q-learning algorithms to set personalized prices conditional on the labels available to them. Because the Q-value update incorporates the maximum Q-value in the next-period segment as a continuation payoff, profits earned in one segment affect the Q-values associated with other segments. We refer to this channel as \textbf{\textit{cross-segment Q-value spillovers}}. 

As a benchmark, we first consider homogeneous segments and symmetric segmentation. In this setting, consumer labels contain no payoff-relevant information, and segmentation merely creates multiple states in the learning environment. We obtain two main findings. First, Q-learning agents tend to sustain collusion through \textbf{\textit{market allocation}} rather than market sharing: each agent monopolizes a subset of segments while competing in the remaining segments. Second, aggregate collusion declines as segmentation becomes finer, in contrast to the classical multimarket-contact prediction of \citet{bernheim1990multimarket}.
We then introduce segment heterogeneity while maintaining symmetric segmentation across firms, so that consumer labels become payoff-relevant. Collusion is \textbf{\textit{unevenly}} distributed across segments: it concentrates in high-value segments, whereas competition intensifies in low-value segments. We further decompose the effect of finer segmentation into two components: an increase in the number of segments and a reduction in uncertainty about consumer type. Both reduce collusion, but the increase in the number of segments accounts for approximately three-quarters of the overall decline.
Finally, we consider heterogeneous segments and asymmetric segmentation. The better-informed agent adopts a \textbf{\textit{Bait-and-Restraint-Exploit strategy}} to induce the less-informed agent to sustain collusive prices. Nevertheless, superior consumer data do not necessarily confer a profit advantage.

Our simulation results shed light on the relationship between data and welfare. First, we show that individual firms' profits does not necessarily increase as their segmentation becomes finer. This reflects that a higher order collusion through coordinating the efficient dataset usage has not been adapted to. We refer to this drawback as ``curse of more data''. Second, firms can strategically select their own data inputs to the AI to maximize profits. We show that the unique Nash Equilibrium results in a dataset profile with the lowest consumer surplus. In this sense, \textbf{\textit{data selection}} will worsen the concern of algorithmic collusion as it can harm consumer surplus even without coordination. Finally, a finer segmentation improves matching efficiency. Meanwhile, it also enables more flexible market allocation schemes, which in turn reduce collusive levels. Therefore, the relationship between industry profits and segmentation refinement is ambiguous. However, the consumer surplus and social welfare generally increase with finer segmentation. 

The implications of our findings span two areas. The first concerns antitrust scrutiny of reinforcement-learning pricing agents, particularly the treatment of market allocation and the use of algorithm audits. The second concerns data policy. Our central point is that many existing policies were formulated without considering their effects on competition, especially that among algorithms. We examine how competition among algorithms is affected by data sharing, data selection, privacy protection, redundant labeling, data advantages, and fairness. We also highlight the potential overuse of consumer data and recommend firms adopt man-made limitation of input dataset.

\paragraph{Related Literature.} Our work contributes to the growing literature on algorithmic collusion. \cite{oecd2017algorithmsandcollusion} raised the concern that AIs may adapt to tacit collusion. This concern has been verified in simulation by \citet{waltman2008q, Calvano2020}, and supported through empirical evidence by \citet{assad2024algorithmic}. One strand of subsequent papers explores the mechanism underlying algorithmic collusion. On the simulation side, \citet{Abada2023,epivent2024algorithmic} show that the seeming collusion may be due to imperfect exploration; \citet{asker2022artificial, asker2023impact} show that asynchronous updating can lead to pricing close to monopoly levels. On the theoretical side, \citet{banchioartificial, bertrand2025self, cartea2025algorithmic} analyze algorithmic pricing dynamics using ordinary differential equation approximations, while \citet{waltman2007theoretical, dolgopolov2024reinforcement, xu2024mechanism} study the problem through the lens of stochastic stability. A complementary strand of the literature uses simulation-based approaches to uncover a rich set of observations across diverse environments, including sequential pricing \citep{klein2021autonomous}, imperfect monitoring \citep{Calvano2023a}, auctions \citep{banchio2022artificial}, platform design \citep{johnson2023platform}, financial markets \citep{colliard2022algorithmic, dou2025ai}. The interaction between rational agents and AIs has been studied by \cite{werner2024algorithmic} and \cite{banerjee2025trading}. \cite{Abada2025} and \cite{Calvano2026} offer a comprehensive survey on this literature. Our work contributes to this literature in three ways. First, our paper simultaneously accounts for both price discrimination and tacit collusion, two important concerns regarding AIs. We examine how consumer datasets affect pricing decisions of AIs in market games. \citet{decarolis2023artificial} also study the role of data in AI-driven competition, but their notion of data refers to opponents' actions (memory versus memoryless learning) and counterfactual estimates (synchronous versus asynchronous learning), whereas ours concerns information about consumer types. Second, we study multi-segment contact between AIs induced by market segmentation, documenting that reinforcement learning algorithms can adapt to tacit collusion through market allocation. Third, to our knowledge, we are the first to study algorithmic collusion among state-based learners and to show that individual learning of the exogenous (uncertain) environment can already coordinate AIs to collude through cross-segment Q-value spillovers.

This study is also situated within the broader literature on price discrimination, information accuracy, and competition (e.g., \citealt{MiklosThal2019,belleflamme2020competitive}). In \cite{MiklosThal2019}, better information is modeled by improving the accuracy of each market while maintaining the number of markets. More accurate information increases the benefit of deviation. Therefore, the collusive prices on the equilibrium path have to be restrained to discourage deviation off the equilibrium path. Our paper studies how price discrimination affects the collusive behavior between AIs. Furthermore, we consider an additional component of finer market segmentation, i.e., an increase in the number of markets. This additional component can further reduce collusion by improving flexibility in market allocation. We also show that this effect outweighs that of uncertainty reduction in market games among AIs.

In addition, our study is related to the literature on multimarket contact among rational, forward-looking decision makers. The fact that multi-market contact can foster anti-competitive outcomes was first proposed by \cite{edwards1955conglomerate}. \cite{bernheim1990multimarket} examined the conditions under which multimarket contact strictly facilitates collusion outcomes in repeated competition. Our work studies multimarket contact between AIs due to market segmentation and finds that AIs can adapt to collusion through market allocation. In addition, we consider uncertainty and asymmetric segmentation. Contrary to the common wisdom in the literature, we show that multimarket contact generally reduces collusive levels in pricing games among AIs. 

Finally, our study connects to the theoretical literature on how correlated signals can coordinate rational decision makers in games. The foundational work by \cite{aumann1974subjectivity,aumann1987correlated} introduced the concept of correlated equilibrium under complete information, which expands upon Nash Equilibrium with correlated signals. \cite{forges1993five} and \cite{bergemann2016bayes} further generalize to games with incomplete information. Our paper shows that correlated signals (as segmentation in the market game) can also affect the coordination among AIs. 

The remainder of the paper is organized as follows. Section~\ref{sec:experiment_design} describes the economic environment and the simulation setup. Section~\ref{sec:complete_information} presents the experimental results for homogeneous segments. Section~\ref{sec:market_segmentation} examines how market allocation schemes evolve under various segmentation scenarios. Section~\ref{sec:implication} discusses the implications, while Section~\ref{sec:robustness_conclusion} provides some robustness checks and concludes the paper.

\section{Experiment Design}\label{sec:experiment_design}
\subsection{Economic Environment}

We consider Bertrand competition among $n$ homogeneous firms, indexed by $i\in\{1,\ldots,n\}$, each with zero marginal cost. Buyers differ in willingness to pay (WTP), denoted by $\omega\in\Omega$, where $\Omega$ is finite. WTP is drawn from a commonly known distribution $F\in\Delta(\Omega)$.

In practice, firms typically do not observe WTP directly. Instead, they use observable buyer characteristics, such as browsing history, device type, location, and demographics, to assign buyers to discrete labels. These labels are informative about WTP only to the extent that the underlying characteristics are correlated with valuations; some labels may be redundant or uninformative. Because firms may use different data or models, the same buyer may receive different labels from different firms. We model these labels as firms' information and refer to them as \textbf{\textit{market segments}} throughout the paper.

Formally, firm $i$ has a prediction technology that induces a \textbf{\textit{market segmentation}} $\langle M_i,\pi_i\rangle$. The set $M_i \subseteq \mathbb{R}$, and $\pi_i(m_i\mid\omega)$ is the probability that a buyer with WTP $\omega$ is assigned to segment $m_i\in M_i$. After observing $m_i$, firm $i$ chooses a price $p_i(m_i)$, where $p_i:M_i\to\mathbb R_+$ is its segment-contingent pricing rule. Let $\mathbf p=(p_1,\ldots,p_n)$ denote the pricing rules and $\mathbf m=(m_1,\ldots,m_n)$ the realized segment profile. Given $(\mathbf p,\mathbf m,\omega)$, define the set of lowest-price firms as
\[
L(\mathbf p,\mathbf m):=\arg\min_{i\in\{1,\ldots,n\}} p_i(m_i).
\]
If $\min_i p_i(m_i)\le \omega$, the buyer chooses one firm from $L(\mathbf p,\mathbf m)$ uniformly at random; otherwise, she does not buy. Firm $i$'s realized demand is
\[
q_i(\mathbf p,\mathbf m,\omega)
=
\mathbf 1\{p_i(m_i)\le \omega\}
\mathbf 1\{i=\xi(L(\mathbf p,\mathbf m))\},
\]
where $\xi(L(\mathbf p,\mathbf m))$ is drawn uniformly from $L(\mathbf p,\mathbf m)$. Firm $i$'s realized profit is
\[
r_i(\mathbf p,\mathbf m,\omega)
=
p_i(m_i)q_i(\mathbf p,\mathbf m,\omega).
\]

\begin{proposition}\label{prop:static_benchmark}
For any market segmentation profile, under any Bayesian Nash Equilibrium, each firm's profit is zero. 
\end{proposition}
\begin{proof}
    The proof of general model can be found in our companion note \cite{xu_market_2026}. 
\end{proof}

Proposition~\ref{prop:static_benchmark} provides the zero-profit competitive benchmark for our simulations. In the simulations, we study repeated interaction in this environment. In each period $t=0,1,2,\ldots$, a new buyer arrives, $\{\omega_t\}_{t\geq 0}$ are independent and identically distributed (i.i.d.) draws from $F$. firms observe their realized segments $m_{i,t}$, and prices are chosen simultaneously. Firms delegate pricing to Q-learning agents, whose policies evolve through reinforcement learning.

Our setting differs from traditional models of market segmentation and multimarket contact, such as \citet{bernheim1990multimarket}, in two respects. First, segmentation is generated by firm-specific data and prediction models, and therefore may be asymmetric across firms. Second, buyers arrive sequentially rather than simultaneously across all markets. Sequential pricing is natural in our environment. Algorithmic pricing is widely used in online settings, where menu costs are negligible and prices can adjust quickly to market conditions. Moreover, each transaction generates new data that can update learning algorithms, making sequential interaction particularly relevant for pricing with learning agents. Simultaneous pricing across many segments would also create a large strategy space, which is difficult to simulate and arguably less plausible in practice. Nevertheless, Section~\ref{subsec:simultaneous_pricing} shows that our main findings are robust to simultaneous pricing.

\subsection{Q-learning}

Within this environment, each firm delegates pricing to a Q-learning agent. Agent $i$ maintains a Q-matrix $\mathbf{Q}_{i,t}$ whose rows correspond to market segments $m_i\in M_i$ and whose columns correspond to feasible prices $p_i\in A_i(m_i)$. The cell $Q_{i,t}(m_i,p_i)$ is the agent's current value of charging price $p_i$ after observing segment $m_i$. In each period, the agent observes the realized segment, selects a price based on the corresponding row of the Q-matrix, observes the resulting payoff and the next segment, and updates the Q-value associated with the visited segment–price pair.

We use Q-learning for three reasons. First, our objective is to draw implications for algorithmic collusion in real-world settings, where future states are typically shaped by both firms' actions and exogenous random variables. As an algorithm designed for general Markov Decision Processes, Q-learning is well suited to model such environments. Second, its simplicity makes the mechanisms underlying collusion more transparent than those generated by more complex reinforcement-learning algorithms. Third, Q-learning is a standard benchmark in the algorithmic-collusion literature, facilitating comparisons with prior work.

\paragraph{Action Selection.}
In period $t$, after observing segment $m_{i,t}$, agent $i$ chooses a price using an $\varepsilon$-greedy rule:
\[
p_{i,t} =
\begin{cases}
\arg\max_{p'\in A_i(m_{i,t})} Q_{i,t}(m_{i,t},p'),
& \text{with probability } 1-\varepsilon_t,\\
\text{a uniformly random price in } A_i(m_{i,t}),
& \text{with probability } \varepsilon_t.
\end{cases}
\]
We set $\varepsilon_t=\exp(-\beta t)$, where $\beta>0$ is the \textbf{\textit{decay rate}} and controls how quickly exploration declines. The $\varepsilon$-greedy rule balances exploration and exploitation. Early in training, agents experiment frequently, which helps them learn the payoff environment from limited initial information. As $\varepsilon_t$ declines, they increasingly exploit what they have learned by choosing prices with high current Q-values.

\paragraph{Updating.}
After prices are chosen, agent $i$ observes its realized payoff $r_{i,t}$ and
the next segment $m_{i,t+1}$. The visited Q-value is updated according to
\[
Q_{i,t+1}(m_{i,t},p_{i,t})
=
(1-\alpha)Q_{i,t}(m_{i,t},p_{i,t})
+
\alpha\left[
r_{i,t}
+
\delta
\max_{p'\in A_i(m_{i,t+1})}
Q_{i,t}(m_{i,t+1},p')
\right], \tag{U}\label{eq:update}
\]
where $\alpha\in(0,1)$ is the \textbf{\textit{learning rate}} and
$\delta\in[0,1)$ is the \textbf{\textit{discount factor}}. All unvisited
segment-action pairs retain their previous Q-values.

One might question the use of the next realized segment in the continuation value, since segments are drawn i.i.d. in the baseline environment and are not affected by firms' prices. We nevertheless use the standard update for three reasons. First, i.i.d. segment realizations do not imply independent payoffs across segments. Data-driven segmentation is firm-specific and need not be observed by rivals. A firm observes only its own segment of a buyer, but does not observe how rivals segment the same buyer or whether they condition prices on the same market segmentation. Thus, rivals' pricing rules can link outcomes across segments. Second, Q-learning is model-free: agents need not know the segment transition process. Third, the standard update nests more general environments with temporally persistent states or state transitions affected by firms' actions; our i.i.d. baseline is merely a simplification.

In this paper, we study coordination through \textbf{\textit{correlating devices}}. Market segments serve as correlating devices, with Q-learning agents conditioning their prices on the realized segments. Relative to \citet{banchioartificial}, our agents are state-based learners that condition their actions on observed states. Relative to \citet{Calvano2020}, our baseline states are exogenous rather than generated by firms' past actions. This design isolates coordination through correlating devices from coordination sustained by repeated-game punishments. Accordingly, the baseline state space excludes past actions; our main findings remain robust when one-period memory is introduced (Section~\ref{sec:robust_memory}).

\subsection{Simulation Setup}\label{sec:simulation design}

We simulate repeated duopoly pricing under each market-segmentation profile. For each profile, we run 1{,}000 independent sessions. In each session, two Q-learning agents interact until the convergence criterion below is satisfied.

\paragraph{Market Segmentation.} 
We consider both redundant and informative market segments. Redundant segments are labels based on payoff-irrelevant characteristics and therefore contain no information about WTP. We isolate this case with homogeneous buyers, setting $\omega=12.5$, the mean WTP in the heterogeneous specification below.\footnote{With homogeneous buyers, the level of WTP does not affect the qualitative results.} This benchmark isolates the effect of the number of segments on competition (Section~\ref{sec:complete_information}). We then consider informative segments. Buyer WTP is drawn uniformly from $\{5,6,\ldots,20\}$.\footnote{We use $\{5,\ldots,20\}$ rather than $\{1,\ldots,16\}$ to avoid a very low bottom valuation relative to the top valuation, which can lead Q-learning agents to ignore low-value segments.}
In this case, market segmentation is an \textbf{\textit{interval partition}} of the WTP support, as shown in Table~\ref{tab:setting}, ranging from a single pooled segment ($|M_i|=1$) to the finest partition in which each WTP level is its own segment ($|M_i|=16$). We study symmetric segmentations, where firms use the same rule (Section~\ref{sec:symmetric simulation results}), and asymmetric segmentations, where firms use different rules (Section~\ref{sec:asymmetric simulation results}). Each section provides the corresponding implementation details.

\paragraph{Action Space.}
Q-learning requires a finite action space, so we discretize feasible prices. We fix the number of actions in each segment at $l$. Let $\overline{\omega}(m)$ denote the highest WTP among buyers in segment $m$. The feasible price set is
\[
A(m)
=
\left\{
\frac{2\overline{\omega}(m)}{l},
\frac{3\overline{\omega}(m)}{l},
\ldots,
\overline{\omega}(m),
\frac{(l+1)\overline{\omega}(m)}{l}
\right\}.
\]
This construction has two features. First, the lowest feasible price, $2\overline{\omega}(m)/l$, is the unique Bertrand-Nash price in segment $m$. Second, the action set includes one price above the maximum WTP, allowing a firm to forgo serving the buyer. When segments differ in WTP, this \textbf{\textit{segment-specific discretization}} keeps the number of actions fixed across segments while allowing the price grid to scale with segment values. The discretization is not \textit{ad hoc}: it avoids mechanically giving high-WTP segments more feasible prices and, when segments are simulated separately, yields nearly identical collusion levels across segments (Online Appendix~B.3). This provides a clean benchmark for studying coordination across segments. Our results are robust to using a common price grid across segments (Online Appendix~D.2).

\paragraph{Baseline Parameters.}
Our baseline simulations consider a duopoly ($n=2$) with an action space of size $l=200$. Following \citet{Calvano2020}, we set $\delta=0.95$ and $\alpha=0.15$. We calibrate a common decay rate $\beta$ to ensure that, under the finest segmentation ($|M_i|=16$), each Q-matrix cell receives approximately 100 random visits in expectation.\footnote{Following \citet{Calvano2020}, let $\nu$ denote the expected number of random visits to a Q-matrix cell over an infinite horizon. With $k=|M_i|$ segments and $l$ actions, $\nu=[kl(1-e^{-\beta})]^{-1}$.}
This yields $\beta=3\times10^{-6}$, which we use for all segmentation profiles. Unless otherwise noted, we use these baseline parameters throughout the simulations. Our main findings are robust to alternative exploration calibrations and parameter choices (Online Appendix~D). Specifically, we recalibrate $\beta$ for each segmentation profile to maintain the same expected number of random visits per Q-matrix cell across different numbers of segments. We also examine robustness to alternative values of $(\alpha,\beta,\delta)$.

\paragraph{Initialization.}
We initialize Q-values as in \citet{Calvano2020}, using the discounted payoff that firm $i$ would obtain if rivals randomized uniformly over feasible prices:
\[
Q_{i,0}(m_i,p_i)
=
\frac{1}{1-\delta}
\mathbb E\left[
r_i(p_i,\mathbf p_{-i},\mathbf m,\omega)
\mid m_i
\right],
\]
where the expectation is over WTP, rivals' segment realizations, rivals' uniformly random prices, and random tie-breaking. Because $\varepsilon_0=1$ and exploration decays slowly, agents initially choose prices almost at random. Thus, for reasonable initial values, early exploration quickly overwrites the initialization, so initial Q-values have little effect on convergent outcomes.

\paragraph{Convergence.}
Multi-agent Q-learning has no general convergence guarantee in strategic environments, so we use a practical stopping rule. A session is classified as converged when, for each firm and each segment, the \textbf{\textit{chosen action}} remains unchanged for 100{,}000 consecutive periods. We impose a maximum horizon of one billion periods. All sessions in our simulations converge under this criterion, with an average convergence time of about five million periods.\footnote{Our criterion differs slightly from \citet{Calvano2020}, who define convergence using stability of the \textbf{\textit{greedy action}}, i.e., the action with the highest Q-value. Our criterion is stricter because it requires the actually chosen action to remain stable in each segment, which ensures that exploration has decayed sufficiently. Together with our low exploration decay rate $\beta$, this conservative criterion makes convergence slower. No session reaches the one-billion-period cap.}

\paragraph{Collusion Index.}
We measure collusion using the \textbf{\textit{collusion index}} (CI) of \citet{Calvano2020}. Let $\bar R$ denote total industry profit at convergence, $R^N$ the static Nash benchmark, and $R^M$ monopoly profit. Define
\[
\operatorname{CI}
=
\frac{\bar R-R^N}{R^M-R^N}.
\]
The index measures the share of monopoly profit achieved above the competitive benchmark. Under asymmetric segmentation, we define $R^M$ as monopoly profit under the finer segmentation; the results are unchanged under a weighted
monopoly benchmark. For segment-level outcomes, define
\[
\operatorname{CI}(m)
=
\frac{\bar R(m)-R^N(m)}{R^M(m)-R^N(m)},
\]
where $\bar R(m)$, $R^N(m)$, and $R^M(m)$ are the corresponding industry profits in segment $m$. In asymmetric-segmentation exercises, $m$ refers to a segment of the coarser firm.

\section{Homogeneous Segments}\label{sec:complete_information}

In this section, we study a benchmark environment with homogeneous buyers and symmetric segmentation. We normalize buyer WTP to $12.5$. Both firms face the same market segmentation with buyers evenly partitioned across $k$ segments, where $k\in\{1,2,4,8,16\}$. This benchmark captures redundant labeling in practice: firms may segment buyers using payoff-irrelevant characteristics or overfitted patterns that contain no information about WTP. It also isolates the effect of increasing the number of segments while holding information fixed, and provides a simple setting for explaining the mechanism behind our main findings.

\subsection{Main Observations}

\textbf{\textit{Market allocation}}, a common form of collusion in practice, involves competing firms dividing consumers or markets among themselves. Firms then either avoid competing in markets allocated to rivals or maintain artificially high prices to discourage entry into those markets. We call segment $m$ \textbf{\textit{exclusive}} to firm $i$ if $p_i(m)<p_j(m)$, so that firm $i$ is the unique low-price firm in that segment. A segment is \textbf{\textit{shared}} if $p_i(m)=p_j(m)$ and demand is split by uniform tie-breaking. Collusion takes the form of market allocation when high prices arise in exclusive segments and market sharing otherwise.

\begin{figure}[!b]
\centering
\includegraphics[width=\textwidth]{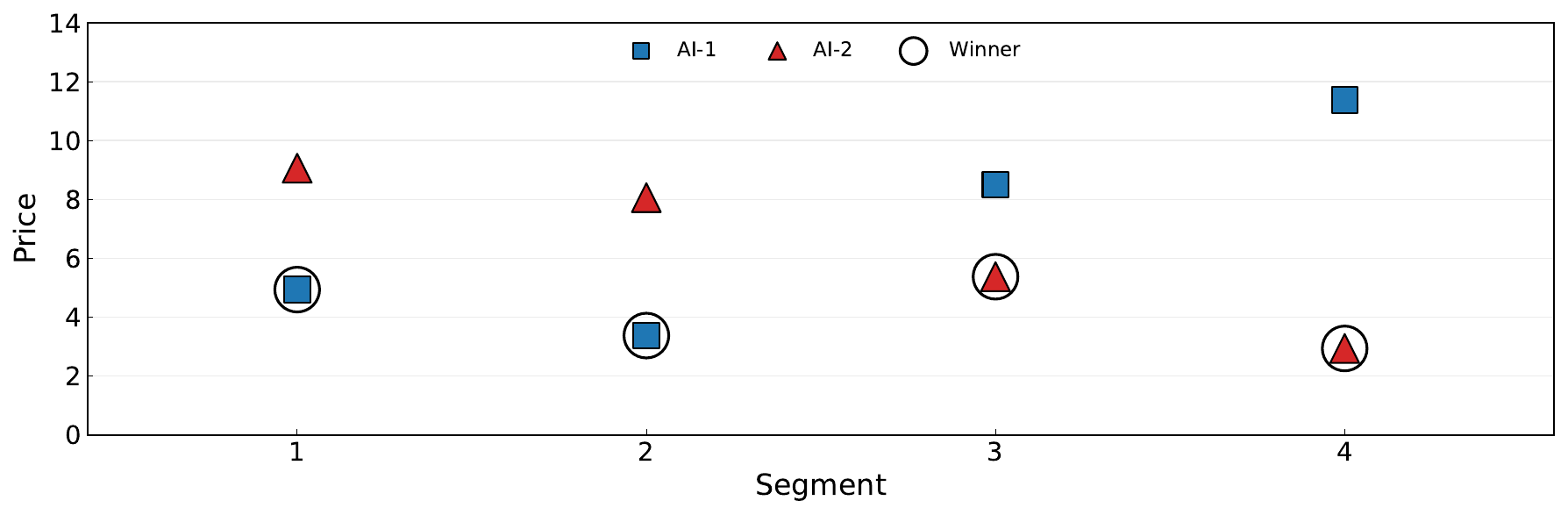}
\caption{Representative Market Allocation across Segments}
\label{fig:market_allocation}
\end{figure}

\begin{obs}[Market Allocation]\label{obs:market_allocation}
In multi-segment environments, market allocation emerges as the dominant mode of collusion: Q-learning agents sustain supra-competitive profits by allocating segments across firms (Figure~\ref{fig:market_allocation}).
\end{obs}

\begin{table}[!t]
\centering
\begin{threeparttable}
\caption{Market Allocation under Homogeneous Segments}
\label{tab:market_allocation}
\small
\begin{tabular}{l ccccc c ccccc}
\toprule
& \multicolumn{5}{c}{Exclusive Segments (\%)} && \multicolumn{5}{c}{Shared Segments (\%)} \\
\cmidrule{2-6} \cmidrule{8-12}
$k$ & 1 & 2 & 4 & 8 & 16 && 1 & 2 & 4 & 8 & 16 \\
\midrule
Total                  & 0.00 & 92.55 & 88.83 & 86.71 & 86.29 && 100.00 & 7.45 & 11.18 & 13.29 & 13.71 \\
$\underline p \ge 25\%\omega$     & 0.00 & 88.20 & 55.23 & 39.85 & 26.92 && 99.70  & 6.90 & 10.65 & 11.93 & 11.79 \\
$\underline p \ge 50\%\omega$     & 0.00 & 46.30 & 26.58 & 17.80 & 11.79 && 76.80  & 5.75 & 8.18  & 7.65  & 7.64  \\
$\underline p \ge \mathbb{E}[\underline p]$         & 0.00 & 42.30 & 38.60 & 31.88 & 24.27 && 49.10  & 7.05 & 9.73  & 11.13 & 11.64 \\
\bottomrule
\end{tabular}

\begin{tablenotes}
\small
\item \textit{Note:} The table reports the share of segments classified as exclusive or shared under homogeneous buyers and symmetric segmentation. ``Total'' includes all segments, regardless of the transaction price $\underline p$. The remaining rows count only segments whose transaction price satisfies the stated threshold: at least $25\%$ of WTP, $50\%$ of WTP, or the average transaction price $\mathbb{E}[\underline p]$ across segments in the corresponding sample.

\end{tablenotes}
\end{threeparttable}
\end{table}

Most studies of algorithmic collusion consider a single market, where collusion must take the form of market sharing (e.g., \citealt{waltman2008q,Calvano2020}). With multiple segments, Q-learning agents instead predominantly coordinate through market allocation. As Figure~\ref{fig:market_allocation} illustrates, each agent becomes the low-price firm in a subset of segments while yielding the remaining segments to its rival.\footnote{\citet{banchioartificial} identify a related collusive mechanism, which they call ``market splitting.'' They study keyword advertising, where advertisers compete in separate second-price auctions for multiple keywords. Their Q-learning agents choose only whether to enter each auction; conditional on entry, bids are automatically set equal to valuations. The environment is a \textbf{\textit{single-state}} learning problem. In contrast, our agents learn segment-contingent pricing rules in a multi-state environment, and market allocation emerges endogenously from learned prices rather than explicit entry decisions.}

Table~\ref{tab:market_allocation} quantifies this pattern. For every $k>1$, more than $86\%$ of segments are exclusive. In the two-segment case, $46.30\%$ of segments are exclusive and have prices above $50\%$ of WTP. The incidence of high-price exclusive segments declines as segmentation becomes finer, but remains substantial: with $k=16$, $26.92\%$ of segments are exclusive with prices above $25\%$ of WTP, more than twice the corresponding share of shared segments.

In repeated pricing games with rational and forward-looking agents, market allocation can be sustained through credible threats to invade the rival's exclusive segments if the rival invades one's own (e.g., \citealt{bernheim1990multimarket}). In contrast, AIs can achieve tacit collusion through market allocation even without one-period memory, which is required to implement trigger strategies. We next examine the subtle differences between market allocation sustained by repeated-game incentives and that arising from algorithmic learning.

\begin{figure}[!t]
    \centering
    \subfigure[Collusion Decline]{\includegraphics[width=0.49\linewidth]{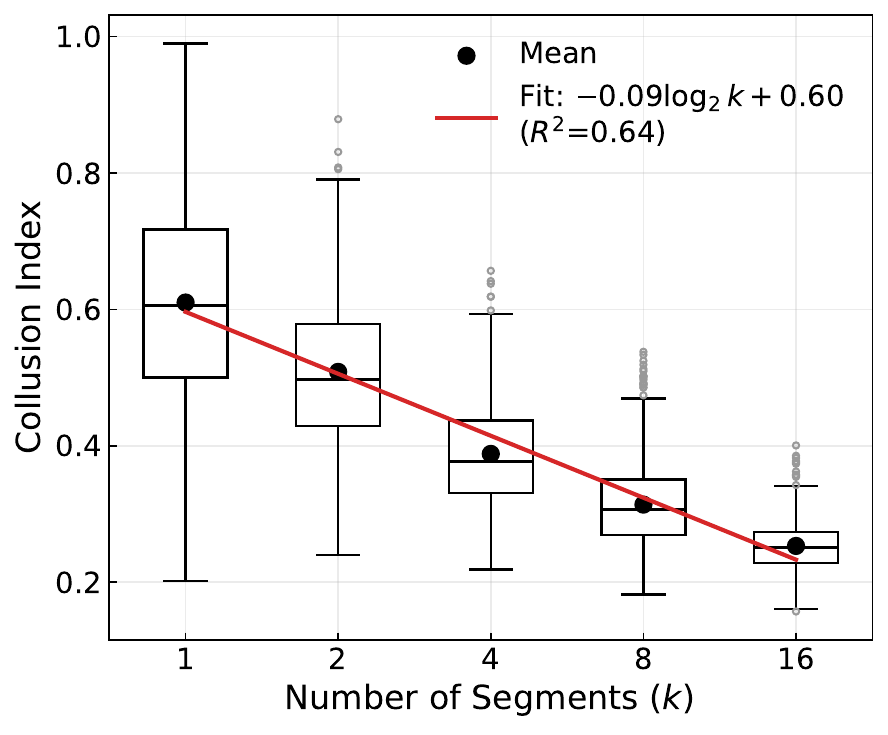}}
    \hfill
    \subfigure[Segment-level CI Correlation, $k=16$]{\includegraphics[width=0.49\linewidth]{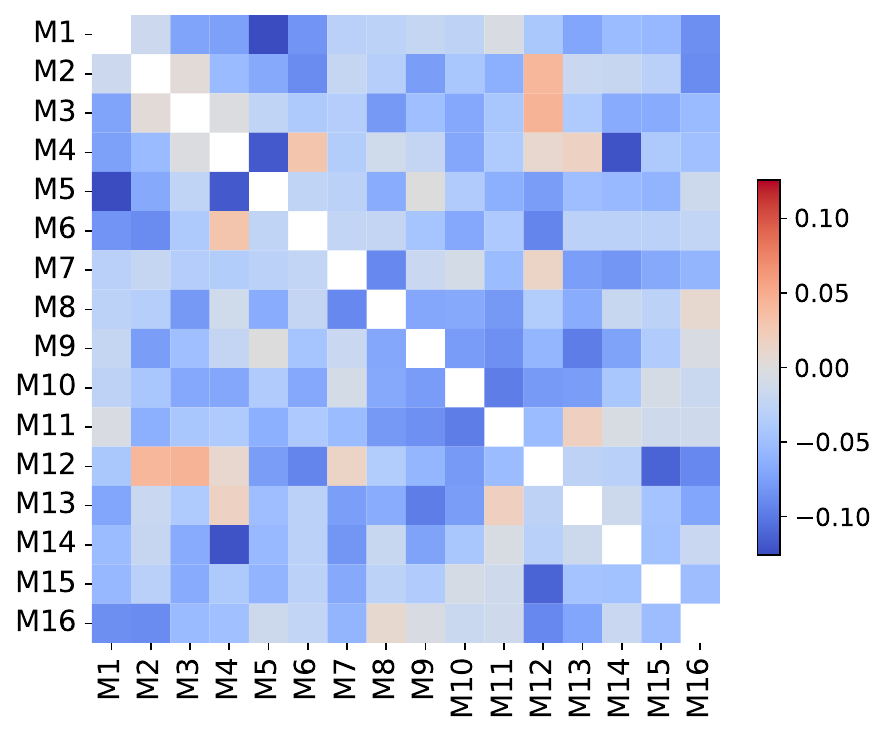}}
    \caption{Declining and Uneven Collusion under Homogeneous Segments}
    \label{fig:collusion_pattern}
    \vspace{2mm}
    \begin{minipage}{0.98\textwidth}
        \small
        \textit{Note:} The left panel shows box plots of aggregate CI by the number of segments; the red line is the regression fit. The right panel reports the pairwise correlation matrix of segment-level CIs across 1{,}000 sessions for the case $k=16$. Blue cells indicate negative correlations, red cells indicate positive correlations, and diagonal entries are omitted.
    \end{minipage}
\end{figure}

\begin{obs}[Declining and Uneven Collusion]
\label{obs:complete_info_profit_corr}
Finer segmentation lowers aggregate CI. Moreover, high supra-competitive profits in some segments are associated with more competitive outcomes in others, generating a negative correlation in segment-level CIs (Figure~\ref{fig:collusion_pattern}).
\end{obs}

This pattern contrasts with standard multimarket-contact logic (e.g., \citealt{bernheim1990multimarket}). With rational firms, finer market segmentation weakly expands the scope for collusion and may strictly increase collusion under heterogeneous segments. Collusion across segments should also move together: higher profits in a firm's exclusive segments strengthen the punishment threat that deters it from undercutting in its rival's exclusive segments. Figures~\ref{fig:collusion_pattern} show the opposite in our learning environment. Under Q-learning, finer segmentation reduces aggregate collusion, even with heterogeneous segments as shown in the next section, and more collusive outcomes in one segment are associated with more competitive outcomes in another.

\subsection{A Brief Explanation}

One key difference between the theoretical results and our simulation findings lies in the contrast between rational, fully informed human reasoning and the behavioral, model-free nature of AI. To account for these observations, we highlight three properties of Q-learning that fundamentally differentiate it from rational human decision-making.

\begin{property}[Experience-Based Decision]\label{prp:action_selection}
    In each segment, the agent chooses the action with the highest Q-value. Whenever an update causes an alternative action's Q-value to exceed that of the current choice, the agent switches to the new maximizer in that 
    segment.
\end{property}

\begin{property}[Asynchronous Update]\label{prp:asynchronous_updating}
    In each period, only the Q-value of the chosen action in the realized segment is updated; all other Q-values remain unchanged.
\end{property}

\begin{property}[Optimistic Bootstrap]\label{prp:bellman_optimality}
    The current update incorporates a Q-value from the segment realized next period, discounted by $\delta$, as the estimate of the continuation payoff. When the exploration rate is small, the next-period action is almost always greedy, so this estimate is effectively the highest next-period Q-value.
\end{property}

To see how the number of segments affects algorithmic collusion, we first examine a single segment. In single-segment Bertrand competition, Q-learning explores the price space and inevitably tests lower prices. Downward exploration captures the entire market, generating a high payoff that reinforces the winner's choice; the loser then keeps switching until it undercuts to a still-lower price. Consistent with common wisdom, this process, which we term \textbf{\textit{alternating undercutting}}, leads to the Nash equilibrium if left uninterrupted.

\begin{figure}[t]
    \centering
    \subfigure[Single-segment Setting (A Sample)]{%
        \label{fig:undercutting_rebound}
        \includegraphics[width=0.49\textwidth]{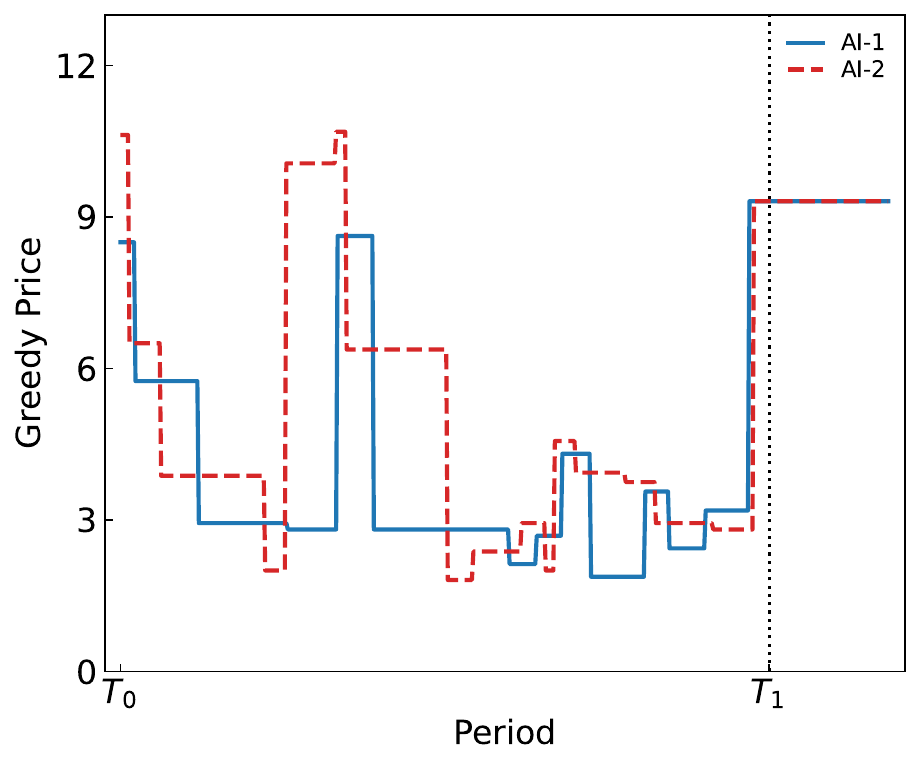}}
    \hfill
    \subfigure[Four-segment Setting (Average)]{%
        \label{fig:last_rebound}
        \includegraphics[width=0.49\textwidth]{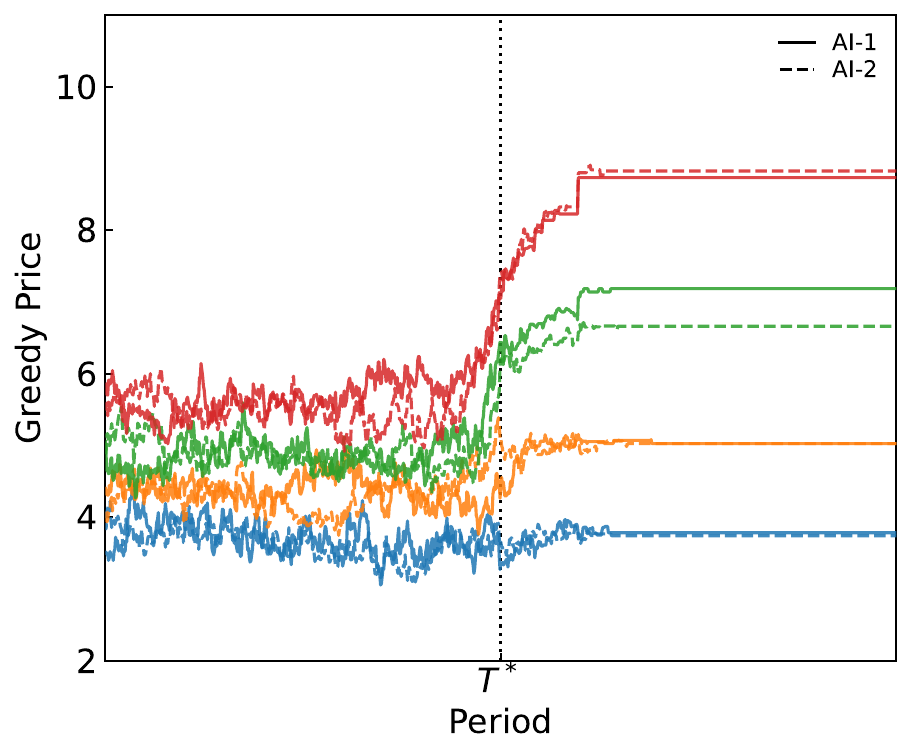}}
    \caption{Alternating Undercutting and Bilateral Rebound}

    \vspace{2mm}
    \begin{minipage}{0.98\textwidth}
        \small \textit{Note:} Panel~(a) presents a representative simulation run in the single-segment setting. The trajectories of AI-1 (solid line) and AI-2 (dashed line) display repeated episodes of alternating undercutting and bilateral rebound, before converging to a high-price outcome that remains stable thereafter. For clarity, any greedy action held for fewer than five consecutive periods is replaced by the most recent sustained greedy action, thereby smoothing out fluctuations. Panel~(b) plots the average price trajectories around the final bilateral rebound at $T^*$ in the four-segment setting, averaged over 1,000 simulation runs, with each color representing a segment.
    \end{minipage}
\end{figure}

Exploration, however, is double-edged: it breaks not only collusion but also,  surprisingly, the alternating undercutting itself. Under exploration, the agents test low prices even when their Q-values are low, and each such update injects the highest next-period Q-value as the estimate of the continuation payoff (Property~\ref{prp:bellman_optimality}). This \textbf{\textit{inflates}} the Q-values of low prices above the profits those prices can sustain. Once a price is undercut sufficiently far, only its Q-value is updated, and it then declines (Property~\ref{prp:asynchronous_updating}), triggering a \textbf{\textit{bilateral rebound}}: both agents jump back to relatively high prices, even when the previous low price is a best response (Property~\ref{prp:action_selection}). Iterating, this process makes the price path resemble an Edgeworth cycle (Figure~\ref{fig:undercutting_rebound}), a cycle also documented by \cite{banchioartificial} in the prisoner's dilemma.\footnote{Our companion paper \cite{xu2024mechanism} documents the same cycle under stochastic discrete updating, rather than the continuous-time approximation of \cite{banchioartificial}, and shows that the price cycle is a short-run phenomenon: a cycle in prices need not be a cycle in Q-values, and in the long run the cycle disappears and play is trapped at the Nash equilibrium. This debate, however, is beyond the scope of the present paper.} As the exploration rate dies out, the two agents may happen to rebound to the same sufficiently high price. Its Q-value then rises sharply while the Q-values of all other actions stay low (Property~\ref{prp:asynchronous_updating}), and the resulting gap deters undercutting for a prolonged period, sustaining supra-competitive prices.

With multiple segments, one critical difference arises: when updating the Q-value of the current-period segment, the algorithm uses the highest Q-value of the next-period segment as the continuation payoff. Consequently, a change in the highest Q-value in one market propagates to the Q-values, and hence the behavior, in other markets. We call this effect \textbf{\textit{cross-segment Q-value spillovers}}. It shapes collusion through both convergence behavior and evolutionary dynamics; here, we focus on the former. In Online Appendix~B.1, we analyze the latter in detail and provide additional supporting evidence.

\paragraph{Market allocation as a new form of collusion.}
A direct implication is the emergence of market allocation (Observation~\ref{obs:market_allocation}). On the one hand, high profits in a firm's self-exclusive segments reinforce its quoted prices by raising the corresponding Q-values. On the other hand, spillovers from these self-exclusive segments into rival-exclusive segments compensate for the low payoffs there, discouraging the firm from undercutting in the segments it has conceded. No explicit punishment is needed: relinquishing some segments becomes self-sustaining through Q-value reinforcement alone. Market allocation is easier to sustain than market sharing, which requires both algorithms to rebound to the \textit{same} high price in the \textit{same} segment; it therefore becomes the dominant form of collusion in multi-segment games.

\paragraph{Collusion in some segments stabilizes competition in others.}
A second implication is a negative correlation in collusive levels across segments (Observation~\ref{obs:complete_info_profit_corr}). Spillovers from collusive segments into competitive ones raise the latter's Q-values and thereby reinforce their competitive prices. Figure~\ref{fig:last_rebound} illustrates this in the four-segment case, plotting the average price dynamics around the final bilateral rebound: the two highest-priced segments rebound clearly, while prices in the remaining two stay low.\footnote{The final bilateral rebound $T^\ast$ is the first period after which the lowest-priced player in each segment remains unchanged. To avoid mistaking transient fluctuations for deviations from convergence, we allow a tolerance of up to 100 subsequent action changes in identifying $T^\ast$.}

\paragraph{Decline of collusion with more segments.}
This mechanism also explains why overall collusion declines as the number of segments increases (Observation~\ref{obs:complete_info_profit_corr}). A bilateral rebound generates large profits that spread through Q-value spillovers, raising the Q-values of other segments above the \textit{threshold} required to sustain competitive prices (Property~\ref{prp:action_selection}). With \textit{few} segments, these spillovers are more \textit{concentrated}, pushing recipient segments far above the threshold. With \textit{many} segments, the same profits are more widely \textit{dispersed}, so each segment exceeds the threshold by only a small margin. Sustaining competition therefore requires a larger fraction of collusive segments when the number of segments is small, while this fraction declines as the number of segments increases.

\section{Heterogeneous Segments and Allocation Scheme}\label{sec:market_segmentation}

In this section, we introduce informative market segmentation by allowing consumers to differ in their WTP. Specifically, WTP is drawn from a discrete uniform distribution over $\{5,6,\ldots,20\}$. We partition this support into $k\in \{1,2,4,8,16\}$ contiguous segments of equal size, as illustrated in Table~\ref{tab:setting}. We begin with the case of symmetric segmentation, in which both firms observe the same market partition, and then turn to environments with asymmetric market segmentation.

\begin{table}[!t]
\centering
\begin{threeparttable}
    \caption{Simulation Setting of Market Segmentation under Heterogeneous Segments}
    \label{tab:setting}

    \small
    \setlength{\tabcolsep}{3pt}

    \begin{tabular}{ccccc}
        \toprule

        $k=16$ & $k=8$ & $k=4$ & $k=2$ & $k=1$ \\
        \midrule

        \multirow{8}{*}{
        \begin{tikzpicture}[
            scale=1.15,
            every node/.style={transform shape}
        ]
            \draw[line width=0.5pt] (0,0) circle (1);

            \foreach \i in {0,...,15}{
                \node[font=\tiny]
                at (22.5*\i+11.25:0.8)
                {\number\numexpr\i+5};
            }

            \foreach \i in {0,...,15}{
                \draw[line width=0.5pt]
                (0,0) -- (22.5*\i:1);

                \node[font=\tiny]
                at (22.5*\i+11.25:1.2)
                {$M{\number\numexpr\i+1}$};
            }
        \end{tikzpicture}
        }
        &
        \multirow{8}{*}{
        \begin{tikzpicture}[
            scale=1.15,
            every node/.style={transform shape}
        ]
            \draw[line width=0.5pt] (0,0) circle (1);

            \foreach \i in {0,...,15}{
                \node[font=\tiny]
                at (22.5*\i+11.25:0.8)
                {\number\numexpr\i+5};

                \draw[
                    dashed,
                    color=gray,
                    line width=0.1pt
                ]
                (0,0) -- (22.5*\i:1.1);
            }

            \foreach \i in {0,...,7}{
                \draw[line width=0.5pt]
                (0,0) -- (45*\i:1);

                \node[font=\tiny]
                at (45*\i+22.5:1.2)
                {$M{\number\numexpr\i+1}$};
            }
        \end{tikzpicture}
        }
        &
        \multirow{8}{*}{
        \begin{tikzpicture}[
            scale=1.15,
            every node/.style={transform shape}
        ]
            \draw[line width=0.5pt] (0,0) circle (1);

            \foreach \i in {0,...,15}{
                \node[font=\tiny]
                at (22.5*\i+11.25:0.8)
                {\number\numexpr\i+5};

                \draw[
                    dashed,
                    color=gray,
                    line width=0.1pt
                ]
                (0,0) -- (22.5*\i:1);
            }

            \foreach \i in {0,...,3}{
                \draw[line width=0.5pt]
                (0,0) -- (90*\i:1);

                \node[font=\tiny]
                at (90*\i+45:1.2)
                {$M{\number\numexpr\i+1}$};
            }
        \end{tikzpicture}
        }
        &
        \multirow{8}{*}{
        \begin{tikzpicture}[
            scale=1.15,
            every node/.style={transform shape}
        ]
            \draw[line width=0.5pt] (0,0) circle (1);

            \foreach \i in {0,...,15}{
                \node[font=\tiny]
                at (22.5*\i+11.25:0.8)
                {\number\numexpr\i+5};

                \draw[
                    dashed,
                    color=gray,
                    line width=0.1pt
                ]
                (0,0) -- (22.5*\i:1);
            }

            \foreach \i in {0,1}{
                \draw[line width=0.5pt]
                (0,0) -- (180*\i:1);

                \node[font=\tiny]
                at (180*\i+90:1.2)
                {$M{\number\numexpr\i+1}$};
            }
        \end{tikzpicture}
        }
        &
        \multirow{8}{*}{
        \begin{tikzpicture}[
            scale=1.15,
            every node/.style={transform shape}
        ]
            \draw[line width=0.5pt] (0,0) circle (1);

            \foreach \i in {0,...,15}{
                \node[font=\tiny]
                at (22.5*\i+11.25:0.8)
                {\number\numexpr\i+5};

                \draw[
                    dashed,
                    color=gray,
                    line width=0.1pt
                ]
                (0,0) -- (22.5*\i:1);
            }

            \node[font=\tiny] at (0:1.2) {$M1$};
        \end{tikzpicture}
        }
        \\

        &&&&\\
        &&&&\\
        &&&&\\
        &&&&\\
        &&&&\\
        &&&&\\
        &&&&\\

        \bottomrule
    \end{tabular}

    \begin{tablenotes}
        \small
        \item \textit{Note}: The table illustrates the simulation design under different levels of market segmentation. Each column corresponds to a segmentation level $k \in \{16, 8, 4, 2, 1\}$. Each circle represents the whole market, and each sector (solid line) represents a segment; the numbers within a sector give the support of consumer WTP in that segment, over which WTP is uniformly distributed.
    \end{tablenotes}
\end{threeparttable}
\end{table}

\subsection{Symmetric Market Segmentation}\label{sec:symmetric simulation results} 

Symmetric market segmentation often arises when the segmentation is induced by public datasets or common tagging systems. For example, e-commerce platforms offer data-driven consumer insights to the competing merchants in their ecosystem: JD offers \textit{JD Shangzhi}, Alibaba provides \textit{Business Advisor}, and Pinduoduo supplies \textit{Duoduo Qingbaotong}. Similarly, ad slot sellers provide user-specific information that helps bidders assess the value of advertising inventory. For instance, Tencent provides the Marketing API 3.0 to allow ad inventory bidders to access its proprietary user data. For the majority of small and medium-sized enterprises, internal capabilities for data acquisition and processing are significantly constrained; consequently, they tend to rely exclusively on external data services provided by third-party platforms. The resulting market segmentation profiles are therefore symmetric.

\begin{figure}[t]
    \centering
    \includegraphics[width=\linewidth]{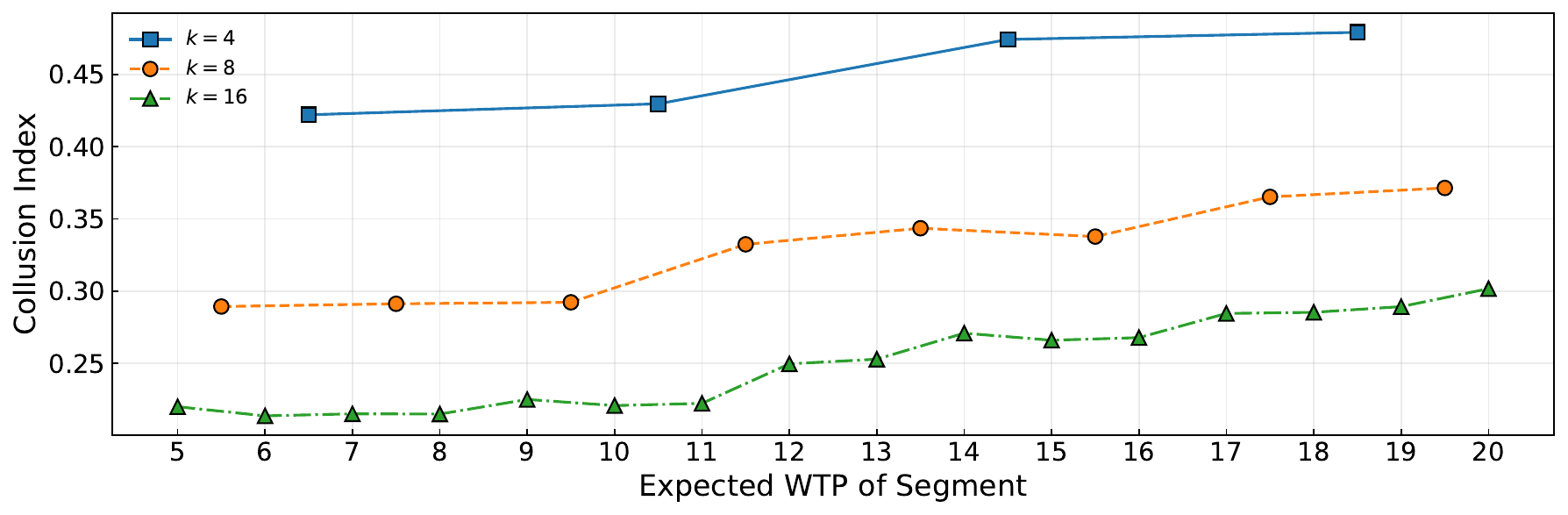}
    \caption{Variation in CI across Segments under Symmetric Market Segmentation}
    \label{fig:collusion_cross_mkt}
\end{figure}

In Section~\ref{sec:complete_information}, we showed that collusion in some segments sustains competitive prices in others through cross-segment Q-value spillovers. This Observation continue to hold with heterogeneous segments. It then raises a natural question: which segments are more prone to collusion? Figure~\ref{fig:collusion_cross_mkt} shows that, once the number of segments reaches at least four, the CI generally rises with a segment's expected WTP.\footnote{In the two-segment case, each player captures one segment in 90\% of the sessions. During the learning process, the two players' Q-values remain close until the rebounds, resulting in relatively small profit differences between them. Because monopoly profits differ substantially across segments ($8.5$ in the low-value segment versus $16.5$ in the high-value segment), the CI tends to \emph{decline} with WTP in this case. See Online Appendix~B.2 for details.}

\begin{obs}\label{obs:ob_CI_WTP}
   Under symmetric market segmentation, segments with higher expected WTP exhibit higher CI when segmentation is sufficiently fine, i.e., $k\geq 4$ (Figure~\ref{fig:collusion_cross_mkt}).
\end{obs} 

Recall that the stability of the entire market-allocation scheme relies on collusive profits in some segments spilling over to sustain competitive profits in the remaining segments. Holding the level of collusive profits fixed, however, sustaining collusion in low-value segments (i.e., segments with lower expected WTP) requires a larger fraction of collusive segments. This more stringent requirement makes such collusion less likely to emerge. In this sense, collusion in high-value segments is more effective at stabilizing the market-allocation scheme as a whole.

Consistent with the homogeneous case, the CI continues to decline as segmentation becomes finer. With heterogeneous consumers, however, finer segmentation has two effects: it increases the number of segments and reduces uncertainty about WTP. This finding is consistent with \citet{colliard2022algorithmic}, who show that reducing uncertainty facilitates the learning of competitive outcomes and thereby weakens collusion. In their framework, the state distribution (buyer WTP in our setting) varies while agents compete within a single signal (a single segment in our setting); in contrast, our framework holds the WTP distribution fixed and varies the segmentation structure. This distinction allows us to disentangle the effects of reduced uncertainty from those of an increased number of segments. Beyond confirming that lower uncertainty mitigates collusion, we show that the increase in the number of segments, which serve as a correlating device through correlated signals, is the dominant force driving the decline in collusion.

\begin{figure}[t]
    \centering
    \includegraphics[width=1\textwidth]{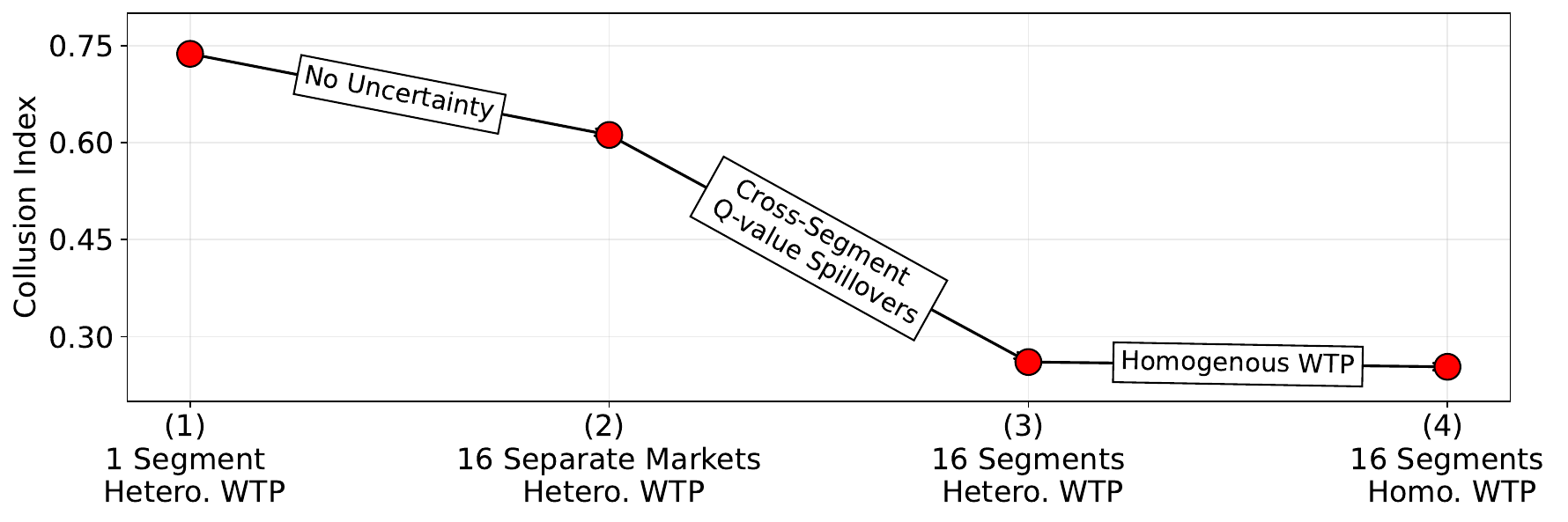}
    \caption{Decomposition of Collusion Decline}
    \label{fig:16_separate_coordinate_11}

    \vspace{2mm}
    \begin{minipage}{0.98\textwidth}
        \small \textit{Note:} This figure compares the CI across the four scenarios described in the main text.
    \end{minipage}
\end{figure}

\begin{obs}\label{obs:primary_factor}
Under symmetric market segmentation with consumer heterogeneity, the decline in the CI as segmentation becomes finer is driven primarily by the increase in the number of segments rather than by the reduction in uncertainty (Figure~\ref{fig:16_separate_coordinate_11}).
\end{obs}

Figure~\ref{fig:16_separate_coordinate_11} compares the CI across four 
scenarios: (1) both firms compete in a single-segment market with 16  uniformly distributed WTP values (Table~\ref{tab:setting}, $k=1$); (2) firms compete in 16 \textbf{\textit{separate}} markets, each corresponding to a single WTP level from 5 to 20, with prices determined by 16 independent Q-learning algorithms, one for each market; (3) firms compete in a 16-segment market with heterogeneous WTP (Table~\ref{tab:setting}, $k=16$); and (4) firms compete in a 16-segment market with homogeneous WTP fixed at $12.5$.

Moving from scenario~(1) to~(2) removes only WTP uncertainty and yields a modest drop in the CI, from $0.737$ to $0.612$. Scenario~(3) additionally enables cross-segment Q-value spillovers, and hence market allocation, producing a sharp drop to $0.261$. Moving from~(3) to~(4) further removes WTP heterogeneity but lowers the CI only slightly, to $0.253$. Thus, as the market segmentation profile changes from $(1,1)$ to $(16,16)$, only $26.3\%$ of the overall decline in collusion can be attributed to the information revealed by segmentation, whereas $73.7\%$ is driven by the increase in the number of segments itself.

\subsection{Asymmetric Market Segmentation}\label{sec:asymmetric simulation results}

This subsection examines a setting with differentiated algorithmic capabilities in market segmentation, which can also be read as firms differing in their  ability to price discriminate. With big data, market segmentation is  typically driven by firm-specific predictive models that estimate individual consumers' WTP. Firms with in-house data capabilities train these models on a combination of self-collected data and datasets purchased from third-party intermediaries. Tencent's Ad Exchange, for example, empowers clients to leverage their proprietary data capabilities during real-time ad decision-making.\footnote{See \url{https://support.e.qq.com/detail?cid=4525&pid=10136}.}
Asymmetries in segmentation then arise from differences in training data, in algorithmic methods, or in both. Larger firms typically hold an advantage, benefiting from exclusive data and greater computational capacity.

Accordingly, we focus on the case where one algorithm produces a strictly finer market segmentation than the other. We refer to the two algorithms as AI-H and AI-L, with AI-H implementing the finer segmentation. We write $(k_H, k_L)$ for the segmentation profile, and set the segments' WTPs as in Table~\ref{tab:setting}, according to the chosen $k_H$ and $k_L$. In what follows, unless otherwise specified, a ``segment'' refers to AI-H's finest partition. We begin with an observation on how AI-H ``teaches'' AI-L to collude.

\begin{figure}[t]
    \centering
    \includegraphics[width=\textwidth]{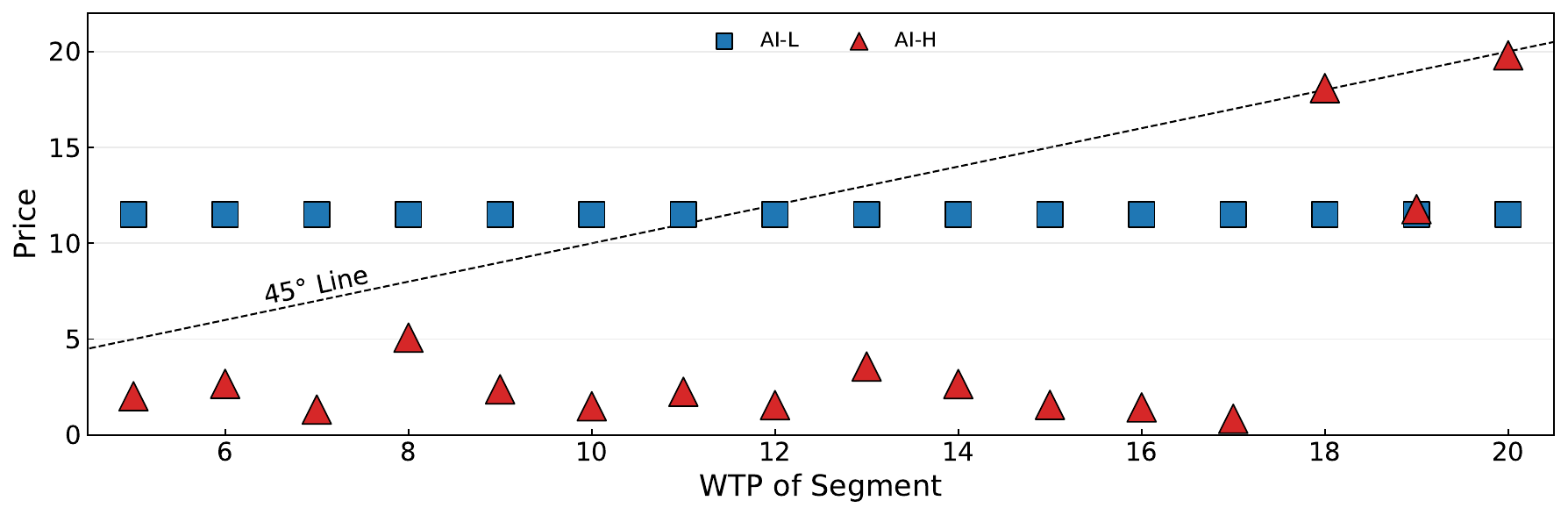}
    \caption{Representative Convergent Prices under Market Segmentation Profile $(16,1)$}
    
    \vspace{2mm}
    \begin{minipage}{0.98\textwidth}
        \small
        \textit{Note:} Red triangles represent the prices set by AI-H across 16 segments, and blue squares represent the uniform price set by AI-L. The dashed line indicates the locus where price equals WTP.
    \end{minipage}
    \label{fig:asym_price_sample_1_16}
    \end{figure}

\begin{obs}[Bait-and-Restraint-Exploit]\label{obs:bait_and_restraint_exploit}
    Under asymmetric market segmentation, AI-H \textbf{baits} in some segments by setting a high price and relinquishing them, while \textbf{restraining its exploitation} in the rest, where it captures demand with a price well below both consumers' WTP and AI-L's quoted price (Figure~\ref{fig:asym_price_sample_1_16}).
\end{obs}

Figure~\ref{fig:asym_price_sample_1_16} illustrates the \textbf{\textit{Bait-and-Restraint-Exploit strategy}} under market segmentation profile $(16,1)$.\footnote{Online Appendix~B.4 reports average results and results for additional segmentation profiles.} Here, AI-H price discriminates across 16 segments, whereas AI-L is restricted to a uniform price. In the segments with WTP equal to 15, 18, 19, and 20, AI-H sets prices strictly above AI-L's uniform price, deliberately forfeiting these segments despite its ability to price discriminate. Although this behavior appears suboptimal, as AI-H could instead undercut AI-L to capture these segments, the bait induces AI-L to maintain a high price. In the remaining segments, by contrast, AI-H engages in restrained exploitation, pricing well below both consumers' WTP and AI-L's price. Table~\ref{tab:bait-and-restraint-exploit} reports the averages: AI-H prices at 68\% of WTP in AI-L's exclusive segments, but at only 34\% in its own exclusive segments.

\begin{table}[!t]

\centering

\begin{threeparttable}
    \caption{Pricing and Deviations under Market Segmentation Profile $(16,1)$}
    \label{tab:bait-and-restraint-exploit}

    \small

    \begin{tabular}{l c cc c ccc} 
    \toprule
    & & \multicolumn{2}{c}{Price/WTP Ratio} & & \multicolumn{3}{c}{Profit Gain from Deviation} \\

    \cmidrule{3-4} \cmidrule{6-8}

    Agent & Profit & AI-H Segments & AI-L Segments & & Upward & Downward & Optimal \\ 

    \midrule

    AI-H & 2.10 & 0.34 & 0.68 & & 2.02 & 2.22 & 4.24 \\

    AI-L & 2.24 & 0.70 & 0.48 & & 0.19 & 0.47 & 0.47 \\ 

    \bottomrule
    \end{tabular}
    
    \begin{tablenotes}
        \small
        \item \textit{Note}: ``Profit'' denotes the average per-period profit. ``Price/WTP Ratio'' is the average ratio of quoted prices to consumers' WTP within the respective exclusive segments of AI-H and AI-L. ``Profit Gain from Deviation'' measures the maximum potential profit increase from unilateral price changes: ``Upward'' (raising price), ``Downward'' (undercutting), and ``Optimal'' (the best possible deviation).
    \end{tablenotes}
\end{threeparttable}
\end{table}

At first glance, as shown in Table~\ref{tab:bait-and-restraint-exploit}, the restrained exploitation strategy appears counterintuitive, as AI-H forgoes up to 2.02 units of profit in its exclusive markets (96\% of its per-period profit) by refraining from raising its price to just below the minimum of AI-L's price and the WTP. However, this strategy serves to reduce AI-L's incentive to undercut. Any local price cut by AI-L would erode its margins in all self-exclusive segments and, at best, yield a modest gain of 0.47 (only 21\% of its per-period profit). Moreover, repeated undercutting would eventually trigger aggressive competition from AI-H. Hence, AI-L is effectively reinforced to maintain the high price and this is why we describe AI-H as ``teaching'' AI-L to collude. The mechanism by which AI-H can learn and maintain this strategy is discussed at the end of this subsection; for now, we focus on its implications for the market allocation scheme and profit sharing.

\begin{figure}[!t]
        \centering
        \subfigure[$(k_H, k_L)= (16,1)$]{\includegraphics[width=0.242\textwidth]{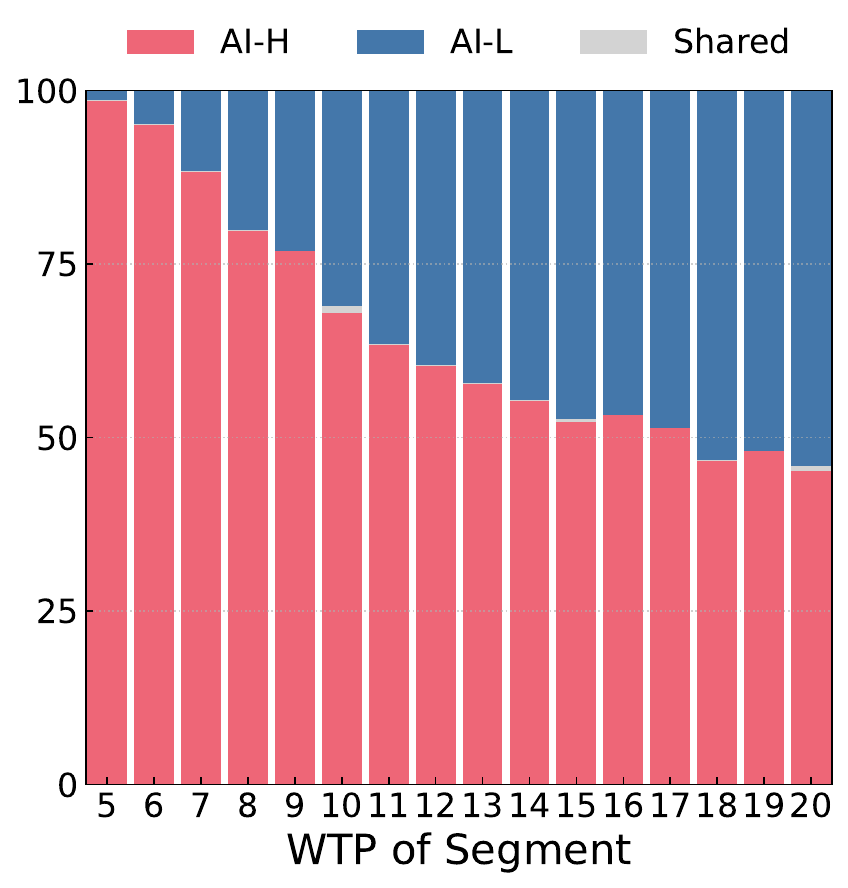}
        \label{fig: market_share_Asym_a}}
        \hfill
        \subfigure[$(k_H,k_L) = (16,2)$]{\includegraphics[width=0.242\textwidth]{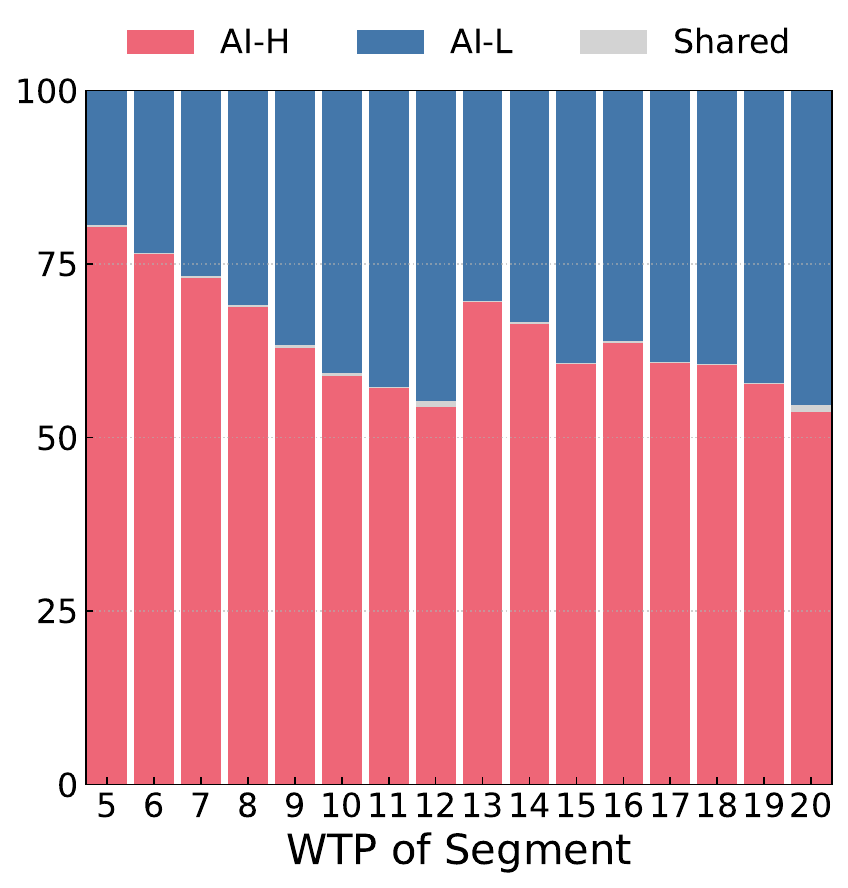}}
        \hfill
        \subfigure[$(k_H,k_L) = (16,4)$]{\includegraphics[width=0.242\textwidth]{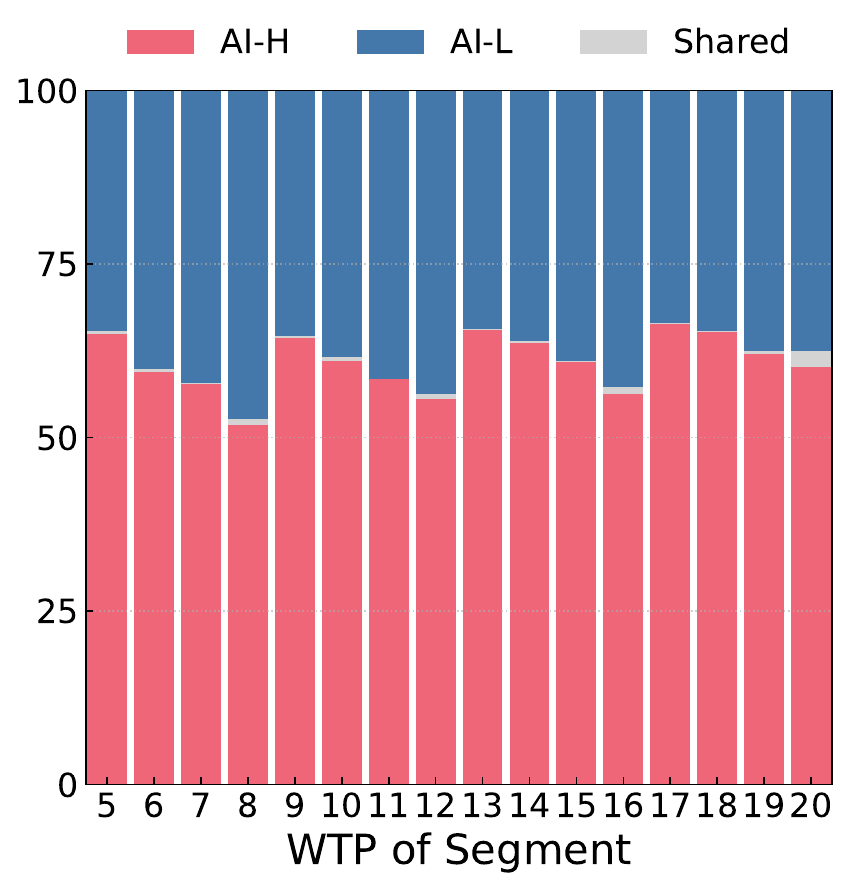}}
        \subfigure[$(k_H,k_L) = (16,8)$]{\includegraphics[width=0.242\textwidth]{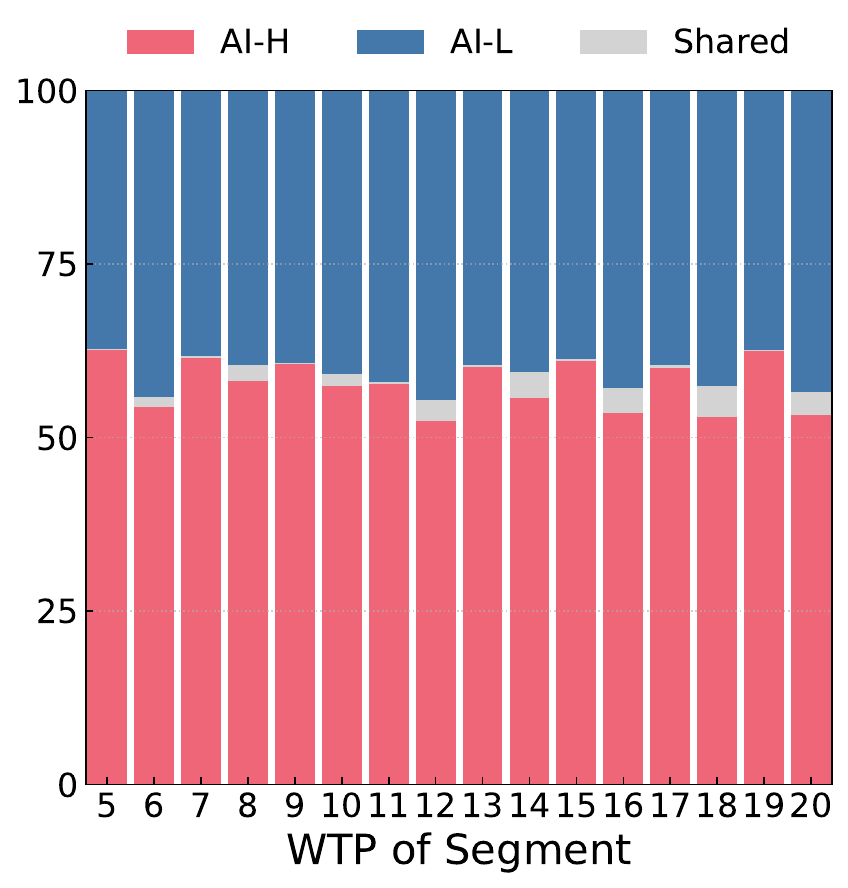}}
        \caption{Market Allocation Scheme under Asymmetric Market Segmentation }
        \vspace{2mm}
        \begin{minipage}{0.98\textwidth}
            \small \textit{Note:} This figure illustrates market allocation between AI-H and AI-L across segments with different consumer WTP levels. Panels (a) through (d) report results for different market segmentation profiles. In each bar, the red, blue, and gray regions represent AI-H's exclusive segments, AI-L's exclusive segments, and shared segments, respectively.
        \end{minipage}
        \label{fig:market_share_Asym}
    \end{figure}

\begin{obs}[Allocation Scheme and Profit Sharing]\label{obs:market_allocation_asym}
    Under asymmetric market segmentation, within each coarser segment defined by AI-L's segmentation, AI-L predominantly serves consumers with higher WTP, whereas AI-H captures consumers with lower WTP (Figure~\ref{fig:market_share_Asym}). Despite this asymmetric allocation of consumers, AI-L earns an average profit comparable to that of AI-H and, in some cases, even exceeds it (Table~\ref{tab:profit_distribution}).
\end{obs}

Figure \ref{fig:market_share_Asym} illustrates market allocation schemes under varying degrees of asymmetric segmentation. Within each coarser market segment defined by AI-L's segmentation, the market share captured by AI-H decreases as consumers' WTP increases. Thus, AI-H strategically yields high-value consumers to its rival as potential bait, while concentrating on capturing low-value consumers. This allocation scheme stems primarily from AI-H's strategy of restrained exploitation. To prevent AI-L from undercutting, AI-H must maintain sufficiently large price gaps between his quoted prices and those of AI-L. As a result, the quantity of segments occupied, rather than their WTP levels, becomes the main determinant of AI-H's profit. Therefore, surrendering a small number of high-value segments is a more efficient way to lure AI-L to maintain relatively high prices.

Table~\ref{tab:profit_distribution} reports the distribution of industry profits, challenging the conventional wisdom that data advantage can translate into competitive advantage in the marketplace. A plausible explanation is that AI-L, which is restricted to offering a uniform price within a coarse market segment, is difficult to collude with. Raising prices forces AI-L to forgo sales to low-value consumers entirely. Consequently, to sustain collusion, AI-H must cede high-value consumers to its rival. At the same time, AI-H must lower its own prices for low-value consumers, thereby limiting its ability to extract consumer surplus. Thus, competitive advantage is sacrificed in exchange for collusion. 

To conclude, we explain how AI-H adapts to this strategic behavior. First, the Q-values implicitly incorporate the opponent's future responses. Specifically, an action taken by AI-H today can alter the opponent's Q-values, thereby shaping its future policy, which in turn influences AI-H's subsequent rewards (Property~\ref{prp:action_selection}). At the same time, each action's Q-value (only) summarizes the estimated discounted payoff of all future interactions triggered by this action (Property~\ref{prp:asynchronous_updating} and \ref{prp:bellman_optimality}). Second, through exploration, AI-H learns that (1) undercutting induces sequential profit losses (due to AI-L’s aggressive responses), and (2) conceding high-value consumers in the current period enables long-term exploitation of low-value consumers. The resulting Q-value updates reinforce AI-H's continued adoption of this strategy. 

One point deserves mention: unlike humans, AI-H is not forward-looking and does not recognize that this exploitation arises from luring its opponent into raising prices. In other words, while AIs can sustain collusion using the Bait-and-Restraint-Exploit strategy, AIs acquires it without understanding its underlying rationale.

\begin{table}[t]
\centering
\begin{threeparttable}
    \caption{Individual Profit across Market Segmentation Profiles}
    \label{tab:profit_distribution}
    
    \small
    \begin{tabular}{l c ccccc} 
    \toprule
    & & \multicolumn{5}{c}{AI-2 ($k_2$)} \\
    \cmidrule{3-7}
    AI-1 ($k_1$) & & 1 & 2 & 4 & 8 & 16 \\ 
    \midrule
    1  & & (2.56, 2.56) & (3.59, 2.44) & \textbf{(3.40, 3.23)} & (2.70, 2.55) & (2.24, 2.10) \\
    2  & & (2.44, 3.59) & (2.89, 2.89) & (3.12, 3.18) & (2.48, 2.51) & (2.06, 2.08) \\
    4  & & \textbf{(3.23, 3.40)} & (3.18, 3.12) & (2.57, 2.57) & (2.26, 2.22) & (1.89, 1.96) \\
    8  & & (2.55, 2.70) & (2.51, 2.48) & (2.22, 2.26) & (2.08, 2.08) & (1.75, 1.79) \\
    16 & & (2.24, 2.10) & (2.08, 2.06) & (1.96, 1.89) & (1.79, 1.75) & (1.68, 1.68) \\ 
    \bottomrule
    \end{tabular}

    \begin{tablenotes}
        \small
        \item \textit{Note}: This table reports the average per-period profit for AI-1 (first entry in parentheses) and AI-2 (second entry) under varying market segmentation profiles $(k_1, k_2)$. Bold entries indicate Nash equilibria in the data selection game.
    \end{tablenotes}
\end{threeparttable}
\end{table}

\subsection{Data and Welfare}
\paragraph{Individual Profit and Data Selection.} Holding the rival's segmentation fixed, individual firm's profit does \textbf{\textit{not}} necessarily increase with the refinement of its segmentation. Its profit even decreases as its segmentation refines, when it has coarser segmentation than the rival does. We term this phenomenon as ``curse of more data''. Note that the curse of more data also happens for rational decision makers in static games (c.f. \citealt{bergemann2016bayes}), since more informed agent cannot commit to relinquish the usage of the additional data while its rivals have perfectly knowledge of its informativeness and therefore change their strategies. For AIs in pricing games, it is well documented that the convergence outcomes deviate from Nash Equilibrium, at least in the short-run. In this sense, a higher order collusion through \textbf{\textit{coordinating data selection}}, well enough to overcome the curse of more data, has not been adapted to.

The curse of more data induces profit-maximizing firms to strategically select the data input to AIs. Since one firm's profit is also affected by the other's dataset selection, a game of dataset selection then emerges. However, as shown in Table~\ref{tab:profit_distribution}, the unique Nash Equilibrium (up to symmetry) is the segmentation profile $(4,1)$. In this case, the industry profit reaches maximum while the consumer surplus reaches minimum. This poses a serious challenge to current regulation. If the tacit algorithmic collusion is not rectified, the dataset selection process will worsen the situation as it can hurt the consumer even without coordination.

    \begin{figure}[t]
        \centering
        \subfigure[Industry Profits]{\includegraphics[width=0.32\textwidth]{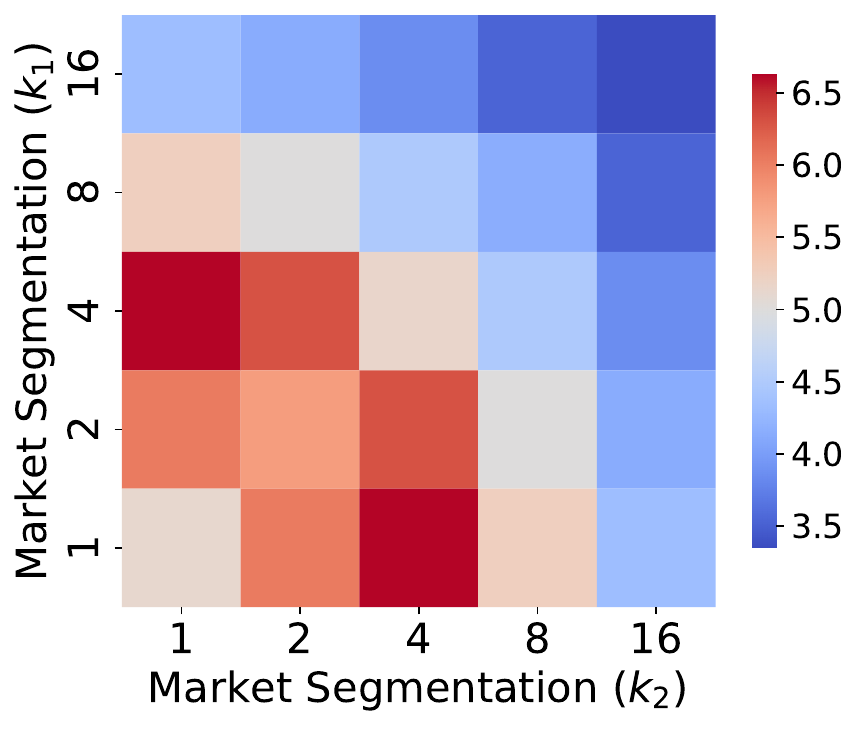}\label{fig:industry_profits}}
        \hfill
        \subfigure[Consumer Surplus]{\includegraphics[width=0.32\textwidth]{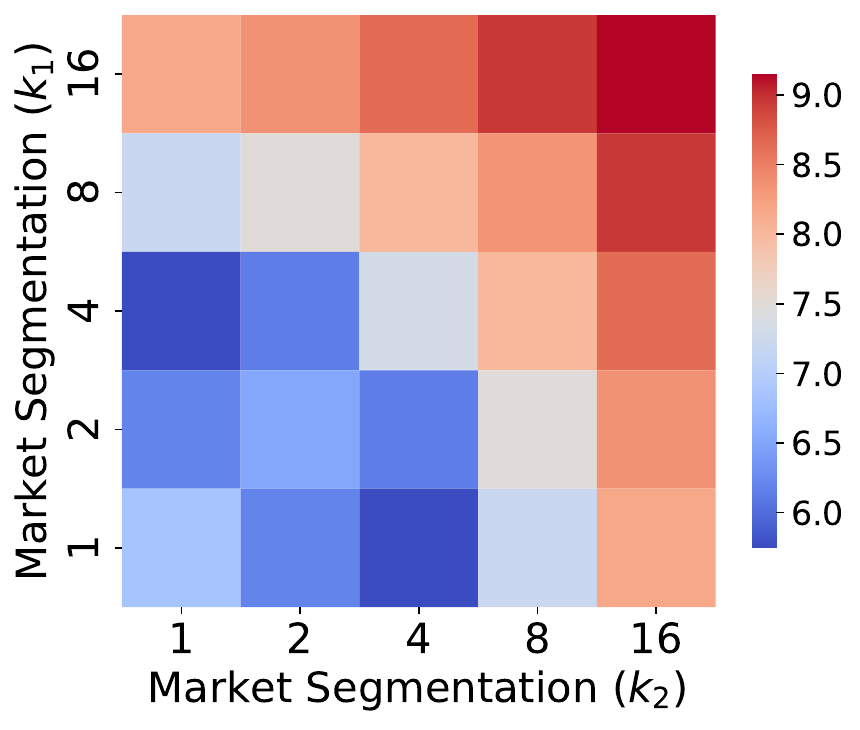}}
        \hfill
        \subfigure[Social Welfare]{\includegraphics[width=0.32\textwidth]{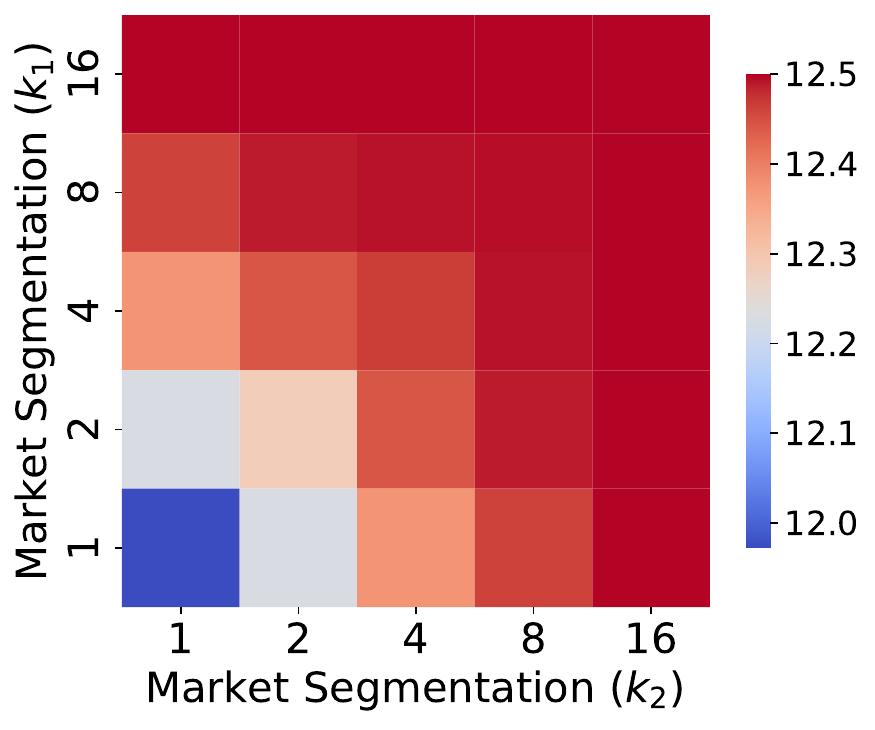}}
        \caption{Welfare cross Market Segmentation Profiles}
        \label{fig:welfare_analysis}
    \end{figure}

\paragraph{Industry Profits.} When both algorithms share the same segmentation, as shown in Figure \ref{fig:industry_profits}, the industry profits exhibit an inverted U-shape as their segmentation refines, which is in line with certain theoretical results (e.g., \citealt{chen2001individual}).\footnote{In \cite{chen2001individual}, firms acquire imperfect information regarding whether consumers are loyal buyers or switchers. They find that industry profits exhibit a U-shaped trend with respect to information accuracy. The upward portion of this trend mirrors our logic: more accurate information improves matching efficiency, thereby boosting industry profits. However, the downward pressure in their framework stems from intensified competition due to better information. This differs fundamentally from our mechanism, where the decline in profits is driven by the decrease of collusion.}  Finer segmentation can improve the matching efficiency but reduce collusive level due to more flexibility in market allocation. The positive effect dominates when segmentation is coarse. As the segmentation refines, the negative effect grows to be dominant. 

When both algorithms have asymmetric segmentation schemes, the effect of refining segmentation of AI-H differs from that of AI-L. Holding AI-H's segmentation fixed, refining AI-L’s segmentation always decreases industry profit. However, holding AI-L's segmentation fixed, the relation between the industry profits and AI-H’s segmentation refinement is not monotonic. The difference lies in the fact that the improved matching efficiency by finer segmentation is absent in the first scenario. 

\paragraph{Consumer Surplus and Social Welfare.}  Figure \ref{fig:welfare_analysis} shows that, both the consumer surplus and social welfare generally benefit from refining firms' segmentation schemes. On one hand, the finer segmentation improves matching efficiency, thereby enhancing social welfare. On the other hand, the finer segmentation reduces collusive level, due to more flexibility in market allocation. Therefore, consumer surplus will increase.

\section{Implications}\label{sec:implication}

Our analysis has implications for antitrust scrutiny in markets where pricing decisions are delegated to reinforcement-learning agents. In particular, it cautions against mechanically extending the per se treatment of explicit market-allocation agreements to market-allocation patterns that emerge endogenously from algorithmic learning. It also highlights new evidentiary and analytical challenges for algorithm audits aimed at assessing competitive effects under a rule-of-reason framework.

\paragraph{Market Allocation and Collusion.} Traditional antitrust enforcement generally regards market allocation as a form of collusion among competing firms (e.g., \citealt{bernheim1990multimarket}) and therefore treats it as a hardcore restriction of competition.\footnote{For instance, the European Commission recently fined Delivery Hero and Glovo \EUR{329} million for, among other infringements, allocating geographic markets to avoid competition. European Commission, ``Commission fines Delivery Hero and Glovo \EUR{329} million for participation in online food delivery cartel,'' Press Release IP/25/1356, July 2025. \url{https://ec.europa.eu/commission/presscorner/detail/en/ip_25_1356}.} 
In contrast, our analysis shows that, when pricing decisions are delegated to competing reinforcement-learning agents, permitting more flexible patterns of market allocation can weaken algorithmic collusion and expand trade, thereby increasing both consumer surplus and total welfare. These findings qualify the conventional antitrust presumption against market allocation in environments governed by autonomous reinforcement-learning pricing.

\paragraph{Algorithm Audit and Antitrust.} Another way to interpret our main results is that individual learning of the exogenous uncertain environment can already coordinate AIs to collude.\footnote{Here we just set the exogenous uncertain environment as i.i.d. to isolate the effect of cross-market spillover.} This poses a new challenge for algorithm auditing as a tool for detecting and addressing algorithmic collusion, as proposed by \citet{Calvano2020science}. In the previous study, to achieve supra-competitive profits in simulation, there are typically two approaches, one through one-step memory, the other through strictly positive discount factor without memory. These two approaches can be detected through auditing algorithms, either through checking whether memory is activated or through checking whether there is a strictly positive discount factor in a static game. When there is uncertainty of exogenous environment, the firm can claim that they design the algorithm just for learning uncertainty without recognizing such practice can collude with other firms.

Next, we draw policy implications on data policy. The main message for them is that most of these policies were made without recognizing its impact on competition, especially competition among algorithms. Our simulation results helps us to re-examine these policies.  

\paragraph{Data Sharing and Efficiency.} Some literature (e.g., \citealt{jones2020nonrivalry}) and current legislative initiatives (like EU Data Act) advocate for data sharing for welfare improvement, due to non-rivalry property of data.\footnote{As articulated in EU 2023/2854, ``The same data may be used and reused for a variety of purposes and to an unlimited degree, without any loss of quality or quantity.'' See \url{https://eur-lex.europa.eu/eli/reg/2023/2854}.} 
Our analysis provides a complementary rationale to support this argument. 
In competitive environments with reinforcement-learning pricing agents, data sharing can facilitate more flexible market-allocation schemes, beyond improving matching efficiency.

\paragraph{Data Selection and Collusion.} Existing regulations for identifying market collusion center on assessing whether firms have engaged in the coordination of prices and output.\footnote{For example, the Lysine Cartel case demonstrates that courts primarily rely on hard evidence of explicit agreements to allocate sales quotas or fix prices to establish liability. For the Department of Justice's official statement on this case, see \url{https://www.justice.gov/archive/opa/pr/1996/Oct96/508at.htm}.} However, our simulation indicates that the choice of consumers' dataset may alter the magnitude of supra-competitive price levels in competitive environments with reinforcement-learning pricing agents. Worse still, in the game of data selection in our setting, the unique Nash Equilibrium (up to symmetry) is the segmentation profile, where the convergent outcome induces the largest industry profit but the least consumer surplus. This then poses a concern since the phase of data selection can drive the outcome toward the worst-case scenario for consumers, even without any requirement of coordination. Therefore, regulators need to closely monitor data input in reinforcement-learning algorithms for each individual firm, especially when algorithmic collusion cannot be avoided completely.  

\paragraph{Privacy and Consumer Surplus.}  To protect consumer privacy, data privacy regulations in major jurisdictions typically enforce ``data minimization'' principles that restrict firms' access to consumer data. Prominent examples include the EU General Data Protection Regulation and China’s Personal Information Protection Law.\footnote{See Regulation (EU) 2016/679, Art. 5.  \url{https://gdpr-info.eu/art-5-gdpr/}. Personal Information Protection Law of the People's Republic of China, Art. 6.  \url{https://personalinformationprotectionlaw.com/PIPL/article-6/}.} However, in the competitive context with reinforcement-learning pricing agents, ``data minimization'' promotes collusion and hurts consumer surplus, via limiting the flexibility in market allocation. Same insight has been drawn in \cite{MiklosThal2019} but with different reasoning.\footnote{In their set-up, better information is modeled by improving the accuracy of each market while maintaining the number of markets. More accurate information improves deviation benefit. Therefore, the collusive prices on equilibrium path have to be refrained to discourage deviation off equilibrium path.} 

\paragraph{Redundant Labeling.} The disclosure of consumer information, though inducing finer market segmentation, may aggravate the privacy concern raised by ``data minimization''. One way to achieve both market segmentation and privacy preserving at the same time is to introduce redundant labels to consumers' dataset for each firm. For instance, the regulator can forcibly assign disparate labels to consumers with similar profiles.

\paragraph{Data Advantage and Fairness.}  Some literature (e.g., \citealt{gans2024market}) and regulators posit that data advantage may result in profit advantage.\footnote{For instance, the Digital Markets Act promotes data interoperability with the explicit aim of ``ensuring for all businesses, contestable and fair markets in the digital sector.'' See Regulation (EU) 2022/1925, \url{https://eur-lex.europa.eu/eli/reg/2022/1925}.} However, in the competitive context with reinforcement-learning pricing agents, the firm with data advantage needs to concede high-valued markets to lure the rival to collude. In this sense, firm with data disadvantage may not necessarily have lower profits. However, more data to firms with data disadvantage, like through data sharing mandated by the Digital Markets Act, may decrease its profit.

\paragraph{Data Overuse and Profit.} Our results also yield a critical managerial implication regarding data usage. In the setting with single-agent decision making, increasing the volume of dataset to an algorithm always improves the performance. However, in competitive environment with reinforcement-learning agents, our findings suggest that more data may result in strictly lower profits. In other words, algorithms fail to collude through coordinating the efficient use of the dataset. Therefore, firms should adopt man-made limitation of the dataset inputted to reinforcement-learning algorithms in this context.

\section{Discussion and Conclusion}\label{sec:robustness_conclusion}

\subsection{Simultaneous Pricing}\label{subsec:simultaneous_pricing}

The standard market segmentation literature assumes that buyers in all markets arrive simultaneously and firms set prices at the same time (e.g., \citealt{bernheim1990multimarket}). In this section, we show that our main results remain qualitatively robust under simultaneous pricing.\footnote{One key difference lies in the tie-breaking rule under simultaneous pricing: when the two firms charge the same price, buyers are evenly split between them. This assumption has little effect on the convergent outcomes when there are more than two segments; see Footnote~4 of the Online Appendix for further discussion.} Further details are provided in Online Appendix~C.1.

\begin{figure}[t]
    \centering
    \subfigure[Collusion Decline]{\includegraphics[width=0.3\textwidth]{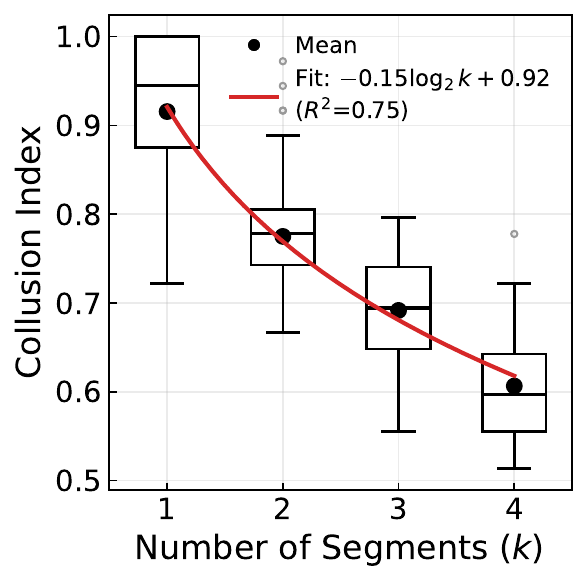}}\hfill
    \subfigure[Segment-level CI Correlation]{\includegraphics[width=0.675\textwidth]{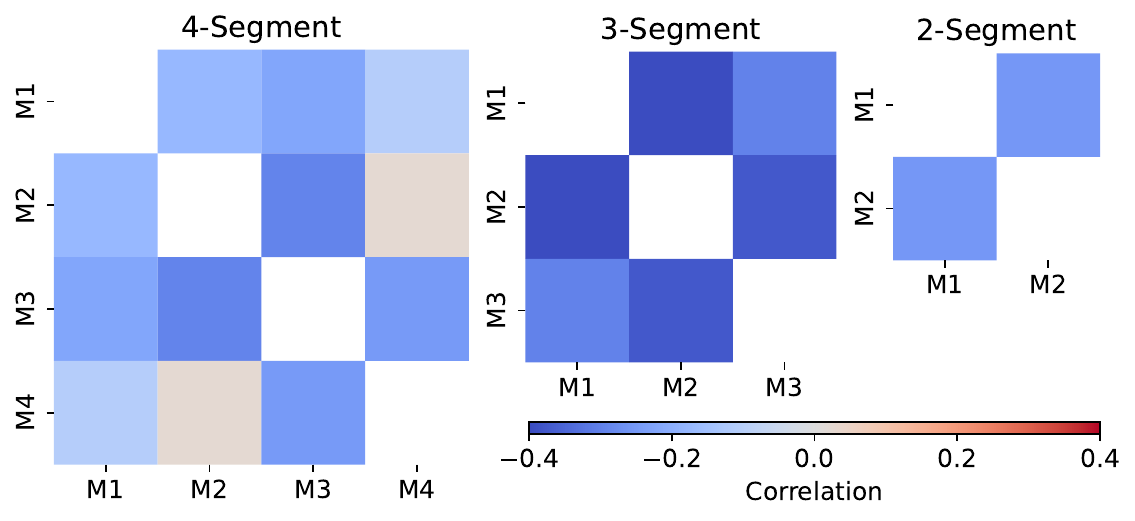}}
    \caption{Declining and Uneven Collusion Under Simultaneous Pricing}
    \label{fig:simultaneous_pricing_collusion}
\end{figure}

Figure~\ref{fig:simultaneous_pricing_collusion} confirms that both collusion decline and the negative correlation remain robust. The cross-segment Q-value spillover effect continues to hold under simultaneous pricing: the bilateral rebound in a subset of segments can generate high profits, which can then reinforce the whole pricing strategy profiles. Simultaneous pricing sustains a higher level of collusion than sequential pricing.\footnote{Under simultaneous pricing, once firms have partitioned the market, each segment generates demand in every period, providing each firm with a stable stream of profits from its assigned segments. Under sequential pricing, by contrast, only one segment is active in each period, and its identity is determined stochastically. Market allocation therefore generates an intermittent and uncertain profit stream, which strengthens firms’ incentives to deviate, particularly when a firm’s assigned segment remains inactive for an extended period.} Furthermore, more than half of the segments are both exclusive and generate profits exceeding half of consumers' WTP. However, with three segments, the allocation proportion reaches its minimum. This is driven by the fact that an odd number of segments precludes a symmetric split, making it difficult to sustain supra-competitive prices solely by market allocation.

\subsection{Q-learning with One-period Memory}\label{sec:robust_memory}

\begin{figure}[t]
    \centering
    \subfigure[Collusion Decline]{\includegraphics[width=0.3\textwidth]{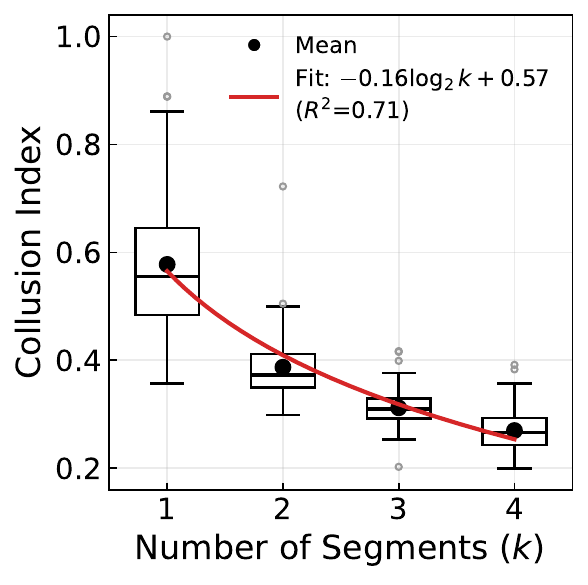}}
    \hfill
    \subfigure[Segment-level CI Correlation]{\includegraphics[width=0.675\textwidth]{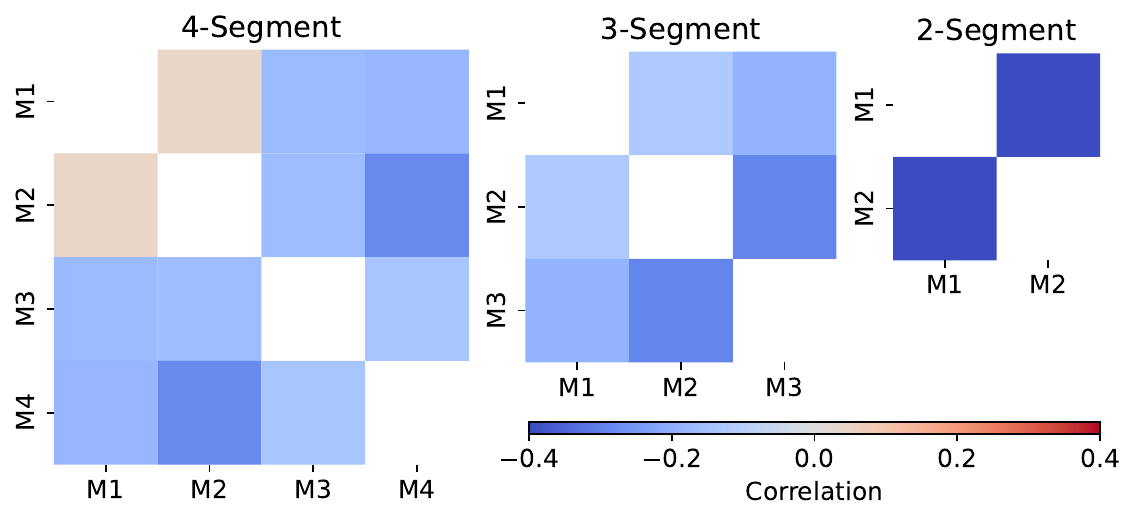}}
    \caption{Declining and Uneven Collusion Under AI with One-period Memory}
    \label{fig:pricing_collusion_memory}
\end{figure}

Following \cite{Calvano2020}, we incorporate one-period memory into the Q-learning algorithm to examine the robustness of our results. Specifically, we allow the algorithms to record the prices set by both firms in the previous period, as well as the specific segment the previous consumer belonged to. We assess the robustness of our results in the baseline setting with homogeneous consumers and symmetric market segmentation. Figure \ref{fig:pricing_collusion_memory} shows that the collusive level of Q-learning algorithms with one-period memory decreases as the number of segments increases. It also that the CIs remain predominantly negatively correlated across segments. Interestingly, the collusive level is much lower than that without memory. Further details are provided in Online Appendix~C.2.

\subsection{Concluding Remarks}\label{sec:conclude}

We study AI-driven algorithmic collusion in markets where competing firms segment consumers into multiple groups. Our results show that AI agents can sustain collusion through tacit market allocation. However, collusion in some segments tends to induce competition in others, generating a negative correlation in collusion levels across segments. Consequently, the overall degree of collusion decreases as the number of segments increases. When markets are heterogeneous, AIs tend to sacrifice low-value segments to competition while sustaining collusion in high-value segments. In cases of asymmetric market segmentation, the AI with finer segmentation adopts a Bait-and-Restraint-Exploit strategy, luring the AI with coarser segmentation into collusion. Yet, data advantages from market segmentation do not necessarily yield higher profits. Finally, the AI with finer market segmentation often ends up serving low-value segments, while the AI with coarser segmentation serves high-value segments. This market allocation scheme is counterintuitive, as one motivation for the purchase of data is to precisely target high-value consumers.

Since our observations hinge on three key properties of Q-learning, namely experience-based decision, asynchronous update, and optimistic bootstrap, we expect our main results to extend beyond Q-learning itself. A broad range of reinforcement learning algorithms, including value-based methods (e.g., Q-learning, SARSA, and DQN), policy-gradient methods, and actor-critic methods, share the first property: actions are adjusted based on past experience, so that an action generating a payoff below its learned expectation becomes less likely to be selected in the future. The other two properties are also present in many reinforcement learning algorithms. Taken together, these three properties imply that reinforcement learning agents, unlike humans, cannot directly recognize best responses. This limitation is precisely what drives our findings and suggests that the observed patterns may arise under a broad range of algorithms. In fact, we have replaced Q-learning with SARSA in Online Appendix~F and found similar patterns.

For future research, three key areas merit attention. First, a more general theoretical analysis of algorithmic collusion is crucial for understanding its underlying mechanisms. Second, studying market segmentation in richer economic environments could provide additional insights. Finally, providing empirical evidence on various algorithmic collusion strategies is essential for gaining a deeper understanding of their dynamics.

\newpage

\bibliographystyle{chicago}
\bibliography{refs}

%% file: appendix.tex
\section{Introduction of Q-learning}\label{app:q_learning}
Q-learning is a fundamental concept in reinforcement learning, a branch of machine learning where an agent learns to make decisions by interacting with an environment. At its core, Q-learning is a model-free, off-policy reinforcement learning algorithm used to find the optimal action-selection policy for a given finite Markov decision process. 

In a stationary Markov decision process, in each period an agent observes a state variable $s \in S$ and then chooses an action $a \in A$.\footnote{In our paper, the ``state'' refers to the identity of the segment from which the consumer arrives, rather than the consumer's specific WTP.} For any $s$ and $a$, the agent obtains a reward $r$, and the system moves to the next state $s^\prime$, according to a time-invariant (and possibly degenerate) probability distribution $P(s^\prime, r|s,a)$. The goal of the agent is to find an optimal strategy $\Gamma^\ast: S\to A$ for choosing actions. A strategy $\Gamma(s)$ is optimal if in each state $s\in S$ it selects an action $a\in A$ that maximizes the agent's cumulative payoff, which is the sum of its immediate payoff and its future payoffs.

Q-learning was proposed by \citet{watkins1989learning} to find an optimal strategy with no prior knowledge of the underlying model, i.e., the distribution $P(s^\prime, r|s,a)$. The algorithm works by iteratively updating an action-value function $Q(s, a)$, which estimates the expected cumulative reward of taking action $a$ in state $s$. The key idea behind Q-learning is the Bellman optimal equation, which states that the optimal action-value function satisfies the equation:
\begin{equation}\label{Q_Bellman}
    Q(s, a) = \sum_{s', r} P(s', r | s, a)\left[r + \delta \max_{a'} Q(s', a')\right],
\end{equation}
where $0 \leq \delta <1$ denotes the discount factor. If the values of the $Q$ function are known, an optimal strategy is given by
\begin{equation}
    \Gamma^*(s) = \arg\max_{a\in A} Q(s,a).
\end{equation}

Q-learning algorithm estimate the $Q$ values iteratively. Starting from an arbitrary initial matrix $\mathbf{Q}_0$, after choosing action $a$ in state $s$, the algorithm observes immediate payoff $r$ and next state $s^\prime$ and updates the estimated $Q$-values, denoted by $\hat{Q}$ using the update rule:
\begin{equation}\label{Q_update_rule}
    \hat{Q}(s, a) \leftarrow (1 - \alpha) \hat{Q}(s, a) + \alpha \left[r + \delta \max_{a^\prime\in A} \hat{Q}(s^\prime, a^\prime)\right],
\end{equation}
where $ 0 < \alpha \leq 1$ is called the learning rate. Note that only for the cell $(s,a)$ visited, the corresponding $Q$-value $Q(s,a)$ is updated.

Exploration and exploitation are two fundamental concepts in reinforcement learning and decision-making processes. To have a chance to approximate the true $Q$ function starting from an arbitrary $\mathbf{Q}_0$, all actions should be tried in all states. Exploration allows the agent to discover the environment and learn about unknown states and actions. Exploitation, on the other hand, involves leveraging the information the agent already has to maximize rewards. Strategies such as $\varepsilon$-greedy, Boltzmann exploration and upper confidence bound are commonly used in various reinforcement learning algorithms and decision-making contexts.

\section{Supplementary Supporting Evidence}

\subsection{Evolutionary Dynamics}\label{app:evolutionary_dynamics}

\begin{figure}[p]
    \centering
    \subfigure{\includegraphics[width=\textwidth]{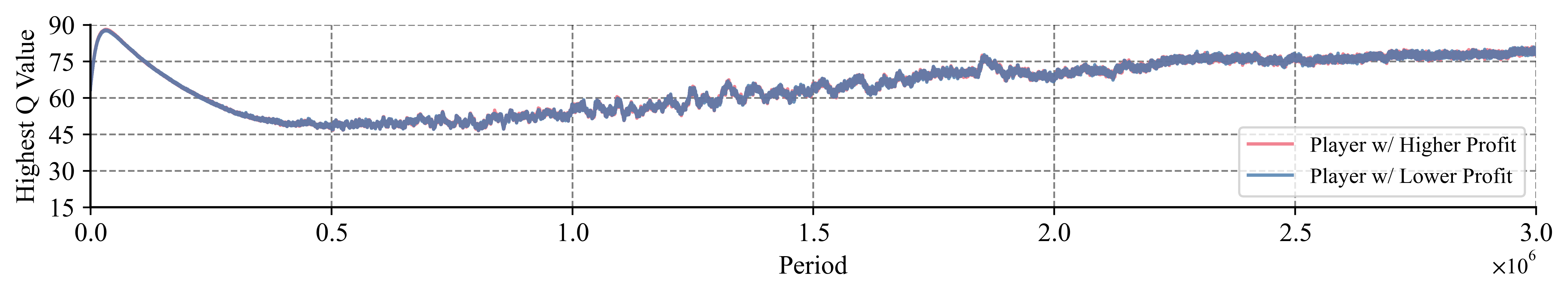}}
    \subfigure{\includegraphics[width=\textwidth]{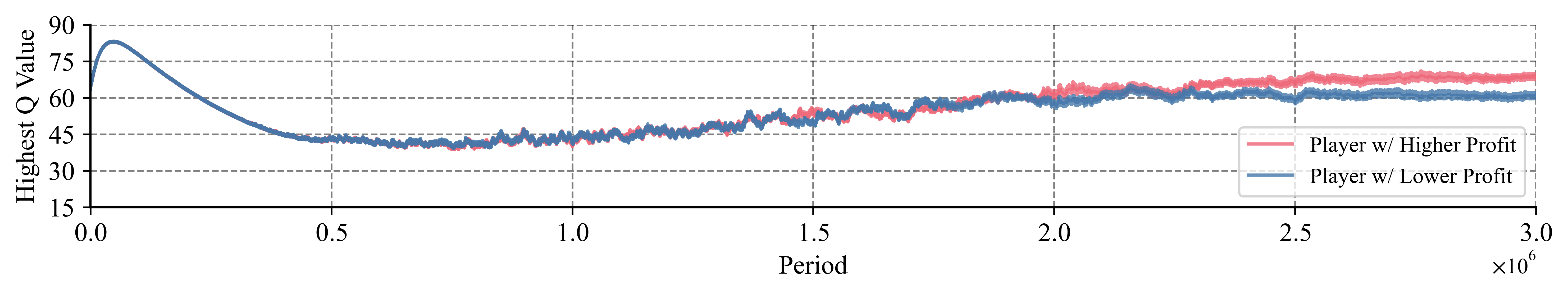}}
    {\includegraphics[width=\textwidth]{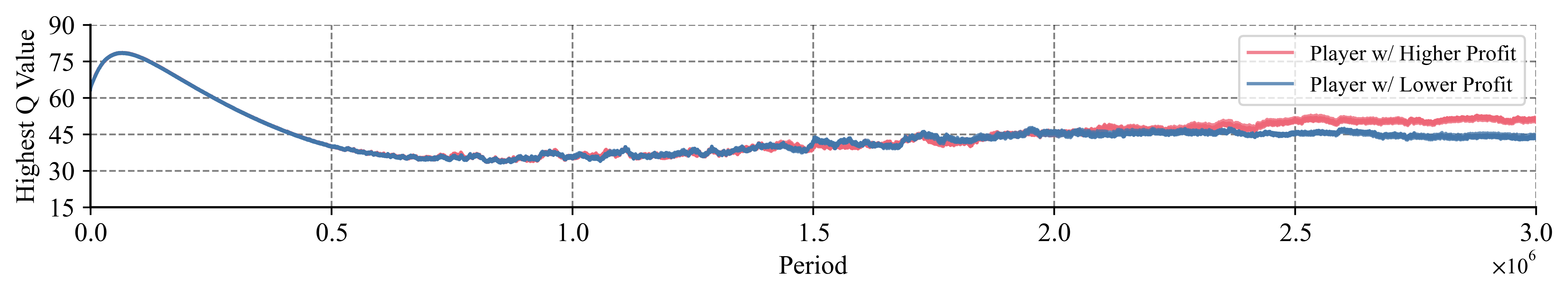}}    
    \subfigure{\includegraphics[width=\textwidth]{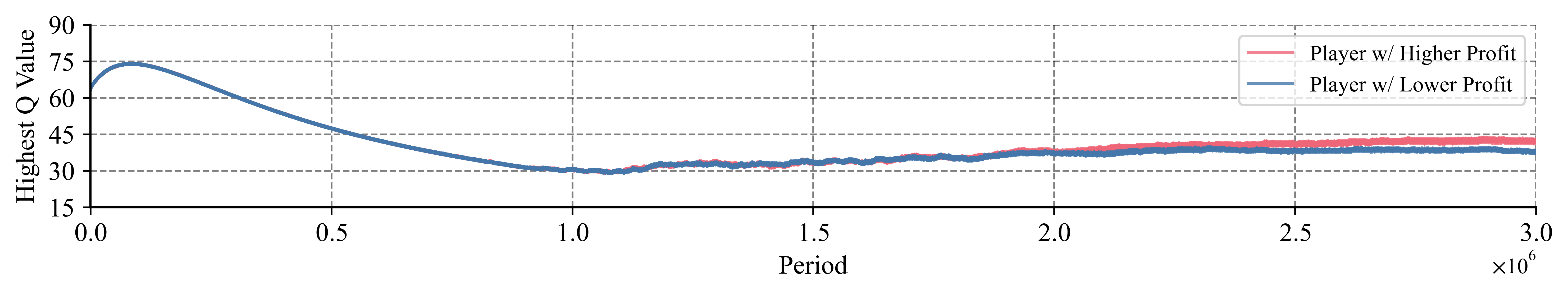}} 
    \subfigure{\includegraphics[width=\textwidth]{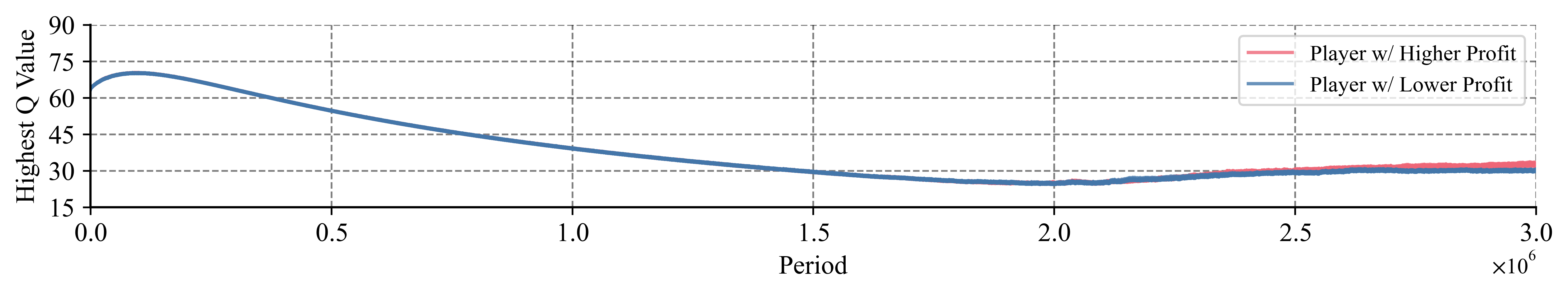}} 
    \caption{Evolution of the Highest Q-Value under Consumer Homogeneity}
    \label{fig:process}

    \vspace{2mm}
    \begin{minipage}{0.98\textwidth}
        \small \textit{Note:}  We track the evolutionary process of the Q-matrix for 100 samples. This figure shows the \emph{average} evolution of the highest Q-value in each segment. From top to bottom, the panels correspond to market segmentation profiles $(1,1)$, $(2,2)$, $(4,4)$, $(8,8)$, and $(16,16)$ in the homogeneous consumer setting.
    \end{minipage}
\end{figure}

In this subsection, we examine the impact of cross-segment Q-value spillovers on evolutionary dynamics of Q-values. Figure \ref{fig:process} illustrates the average evolution of the highest Q-values of each segment, in the benchmark settings. In all cases, the Q-values follow a two-phase pattern: an initial decline followed by a recovery. As the number of segments increases, (i) recovery magnitudes become smaller and more gradual; and (ii) Q-values decline more slowly, with longer downward phases and reaching lower minimum values. These patterns help explain why having more segments leads to lower profits.

First, in the recovery phase, bilateral rebound only leads to smoother and weaker rise in Q-values, as the number of segments increases. This argument is based up by two reasons. With more segments, as argued in the main text, on one hand, the proportion of segments with bilateral rebounds decreases in expectation; on the other hand, the profits generated in these segments with bilateral rebound are distributed across all segments more dispersively. To support this argument, Table~\ref{tab:rebounds_counts} reports the number of rebounds in the highest Q-value series. Over 500 periods, the number of sharp Q-value increases (by 5) declines dramatically from 31,107 in the single-segment case to 234 in the 16-segment case. Within the first $10^6$ periods, which capture the early stage of Q-value adjustment, the 16-segment case exhibits only two such rebounds.

Second, more segments makes it harder to be stable, leading to more sufficient price undercutting process in the decline phase. To see this, consider what happens when one segment happens to experience a bilateral rebound. The high profit generated in this segment is diluted as the number of segments increases, which has two implications. First, the minor increase in Q-values of the rest segments can not interrupt the ongoing price undercutting process. Second, the weak recovery in the rebounding segment makes the recovery phase fragile and easily reversed. This then results in falling back into price undercutting shortly thereafter. Only when Q-values have fallen sufficiently low does it become feasible for just a few segments to rebound and sustain themselves, stabilizing mild collusive outcome.

\begin{table}[!h]
\centering
\begin{threeparttable}
    \caption{Rebound Counts of the Highest Q-values under Consumer Homogeneity}
    \label{tab:rebounds_counts}
    \small

    \begin{tabular}{l ccc c ccc c ccc} 
    \toprule
    & \multicolumn{3}{c}{$(T, \Delta) = (500,5)$} & & \multicolumn{3}{c}{$(T, \Delta) = (500,10)$} & & \multicolumn{3}{c}{$(T, \Delta) = (1000,10)$} \\[2pt]
    \cmidrule{2-4} \cmidrule{6-8} \cmidrule{10-12}
    Seg. ($k$) & $\le 10^6$ & $\le 2\times 10^6$ & Total & & $\le 10^6$ & $\le 2\times 10^6$ & Total & & $\le 10^6$ & $\le 2\times 10^6$ & Total \\
    \midrule
    1  & 2193 & 10146 & 31107 & & 490  & 2518 & 7835 & & 491  & 2527 & 7865 \\
    2  & 589  & 2805  & 9549  & & 100  & 455  & 1353 & & 107  & 505  & 1571 \\
    4  & 156  & 823   & 3127  & & 15.8 & 66.9 & 140  & & 19.9 & 91.4 & 237  \\
    8  & 20.7 & 182   & 844   & & 0.26 & 4.54 & 10.2 & & 0.42 & 9.19 & 21.6 \\
    16 & 2.05 & 31.6  & 234   & & 0.00 & 0.14 & 1.00 & & 0.00 & 0.18 & 2.06 \\
    \bottomrule
    \end{tabular}

    \begin{tablenotes}
        \small
        \item \textit{Note}: A rebound at $t$ with $(T, \Delta)$ is defined as an increase in the Q-value of no less than $\Delta$ within the period $[t-T, t]$. To avoid double-counting, we define $t^{-}_k := \max \{t_k-T, t_{k-1}\}$ and identify the $k$-th rebound at time $t_k$ if $Q_{t_k} - \min_{t' \in [t^{-}_k, t_k]} Q_{t'} \ge \Delta$. Rebounds are counted separately for the highest Q-value series of each segment and player, with reported values representing the grand average across all players, segments, and simulation sessions. In this table, we use $(T, \Delta) = (500, 5)$, $(500, 10)$, and $(1000, 10)$ and report averages for the first $10^6$ periods ($\leq 10^6$), the first $2\times 10^6$  periods ($\leq 2 \times 10^6$), and the entire time horizon (Total).
    \end{tablenotes}
\end{threeparttable}
\end{table}

\subsection{Last Rebound Dynamics of Two-Segment Settings}\label{app:last_rebound_2_mkt}

\begin{figure}[!h]
    \centering
    \subfigure[Market Sharing (10 Samples)]{\includegraphics[width=0.49\textwidth]{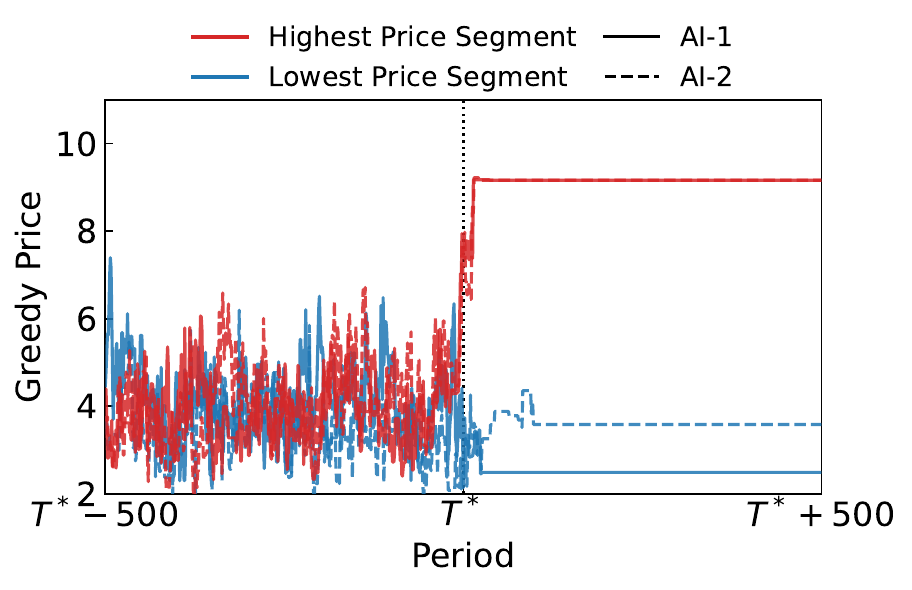}}
    \hfill
    \subfigure[Market Allocation (90 Samples)]{\includegraphics[width=0.49\textwidth]{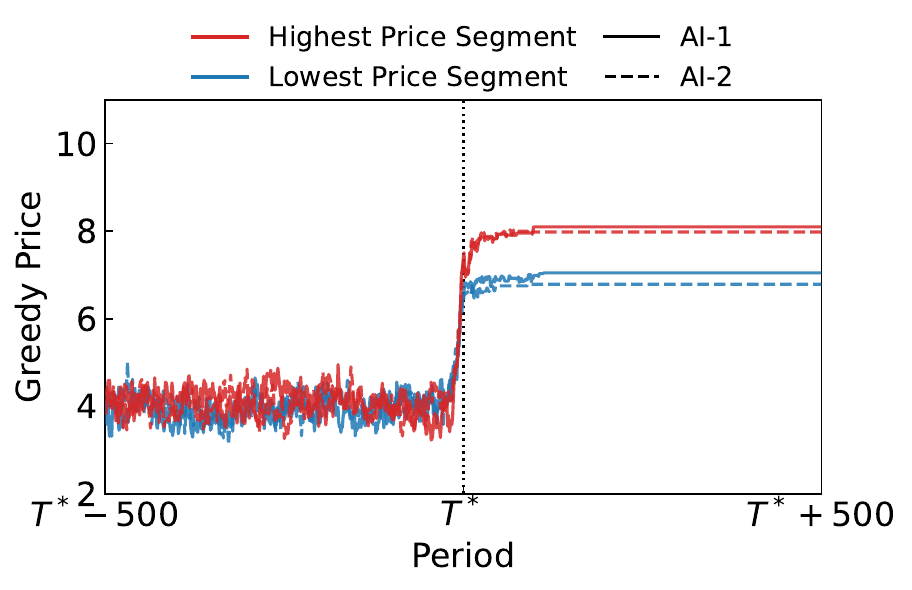}}
    \caption{Last Rebound Dynamics of Two Outcome Types in Two Homogeneous Segments}
    \vspace{2mm}
    \begin{minipage}{0.98\textwidth}
        \small \textit{Note:} The left panel illustrates the rebound strategy under a market-sharing scheme, where both players simultaneously revert to a high price in the same market segment (red line) while one player obtains the other market by setting a sufficiently low price. The right panel illustrates a market-allocation pattern in which each agent serves primarily one market segment while setting a relatively high price in the segment allocated to its rival. The figure reports prices averaged across 90 simulation runs. Consequently, although market allocation emerges in individual runs, AI-2’s average price is lower than AI-1’s in both segments.
    \end{minipage}
    \label{fig:last_rebound_2signal}
\end{figure}

In a two-segment environment, we classify convergent outcomes into two types. The first is the \textit{market-sharing} outcome, where both players share the segment with the highest transaction price. The second is the \textit{market-allocating} outcome, where each player exclusively occupies one segment. Figure~\ref{fig:last_rebound_2signal} illustrates the last rebound dynamics of both types.

In the \textit{market-sharing} outcome, a rebound to a high price in one segment is typically accompanied by persistently low prices in the other---consistent with Section~3, where collusion in some segments induces competition in the rest. In contrast, the \textit{market-allocating} outcome sustains high prices in both segments. Market-sharing is relatively rare, occurring in only about 10\% of two-segment cases. This is because simultaneous rebounds to high prices in both segments, which could be different, are more likely than both players rebounding to the same high price in a single segment.

In the two-segment case, the CI tends to be smaller in the segment with higher expected WTP, inconsistent with Observation~3. This inconsistency arises from two distinct mechanisms. First, with only two segments, market allocation, as the predominant mode of collusion, always assigns both segments for collusion. Therefore, the main mechanism underlying Observation~3, i.e. bilateral rebounds in high-value segments is more efficient in reinforcing competitive prices in the other segments, is missing. However, the mechanism still holds in market-sharing outcome.\footnote{The negative correlation of CI is mainly driven by market sharing outcomes. If we exclude these outcomes, the CI will be positively correlated.} Observation~3 still holds if restricting to this outcome category. Second, the substantial disparity in monopoly profits ($8.5$ in the low-value segment versus $16.5$ in the high-value segment) makes the CI in the low-value segment more likely higher.

\subsection{Collusion under Heterogeneous but Separate Segments}
\label{app:hetero_separate_mkt}

Remind that cross-segment spillover can reinforce competitive price levels and therefore decreases collusive level. To support this fact, we design an auxiliary experiment with heterogeneous consumers divided into separated segments. For each firm, there is an independent Q-learning algorithms assigned to each segment, whose objective is to maximize profits in their assigned segments alone. Simulations in each separated segment are run independently $100$ times. We then aggregate the results to compute total profits and the CI. By design, this decoupled structure precludes the cross-segment spillover effect.

\begin{figure}[!h]
\centering

\includegraphics[width=0.65\textwidth]{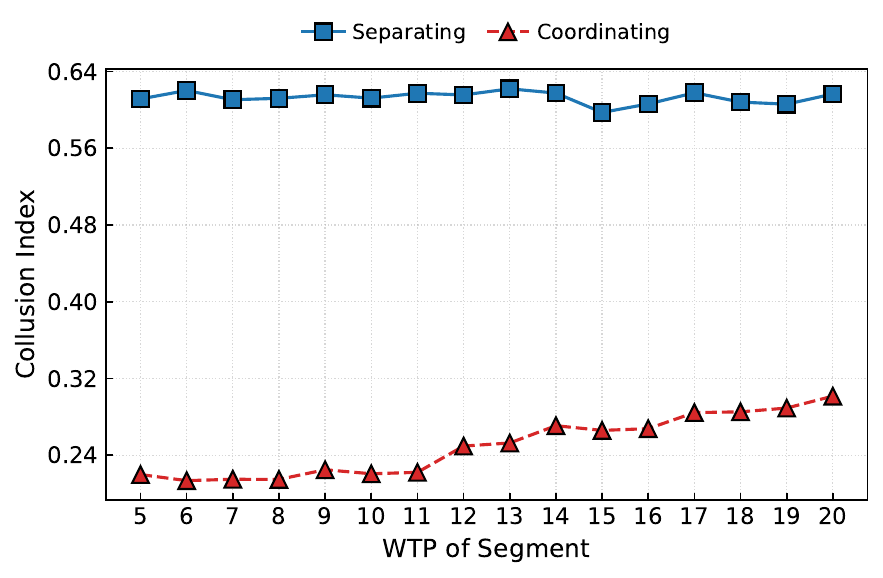}
\caption{CIs across Markets under Market Segmentation Profile $(16,16)$}
    \vspace{2mm}
    \begin{minipage}{0.98\textwidth}
        \small \textit{Note:} The blue squares represent the average CIs of the 16 individual segments, while the red triangles represent the average CIs under the heterogeneous-consumer $(16,16)$ market-segmentation profile.
    \end{minipage}
\label{fig:CI_separate_mkt}
\end{figure}

Figure~\ref{fig:CI_separate_mkt} compares the CI of each segment under the separate segment setting and the market segmentation. First, in the setting of separate segments, each segment converges to the same CI value, which contrasts with the Observation~3. Second, the collusive level is significantly higher in the setting of separate segments. These two phenomenon supports the argument that the interaction among different segments through cross-segment spillover can reinforce competitive prices in low-value segments and reduce overall collusion degree. 

\subsection{Statistics about Bait-and-Restraint-Exploit Strategy}
\label{app:bait_and_restrained_exploit}

\begin{figure}[!h]
    \centering
    \subfigure[$(k_H,k_L)=(16,1)$]{\includegraphics[width=0.49\textwidth]{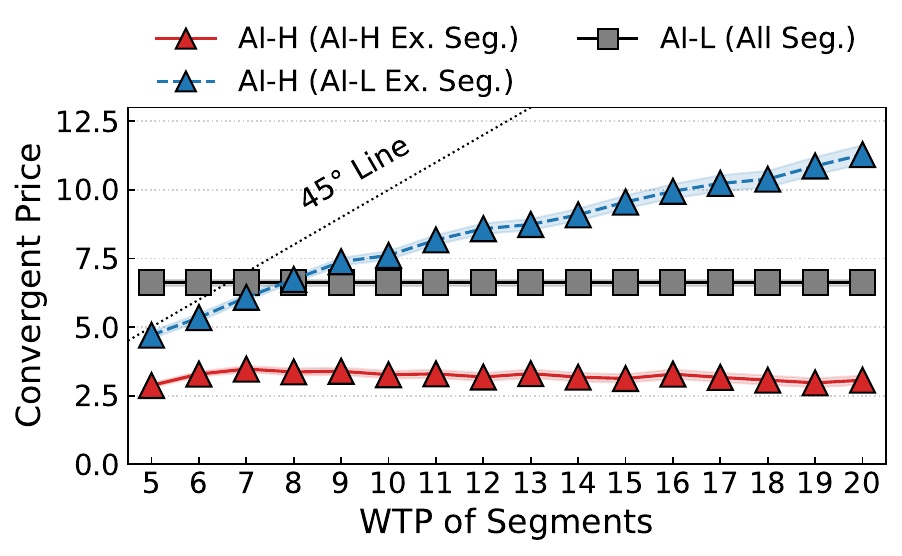}}
    \hfill
    \subfigure[$(k_H,k_L)=(16,2)$]{\includegraphics[width=0.49\textwidth]{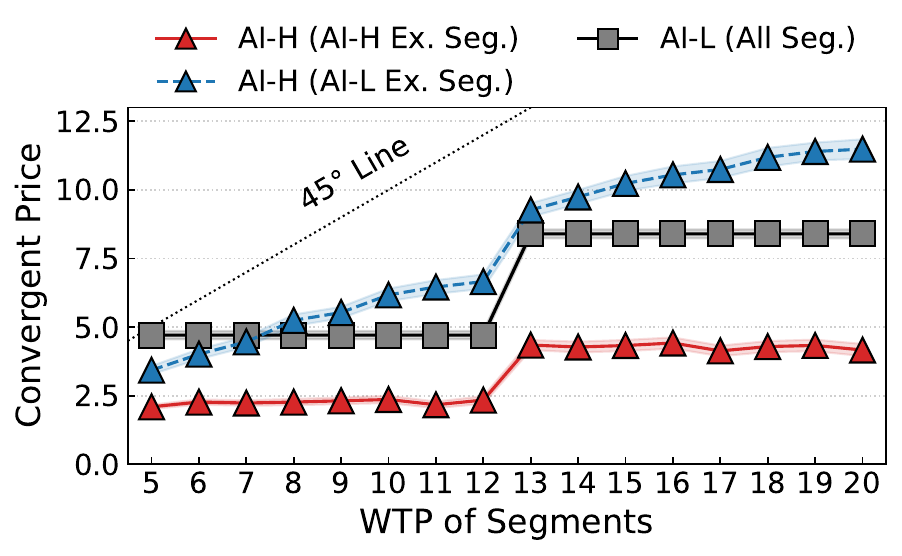}}
    \hfill
    \subfigure[$(k_H,k_L)=(16,4)$]{\includegraphics[width=0.49\textwidth]{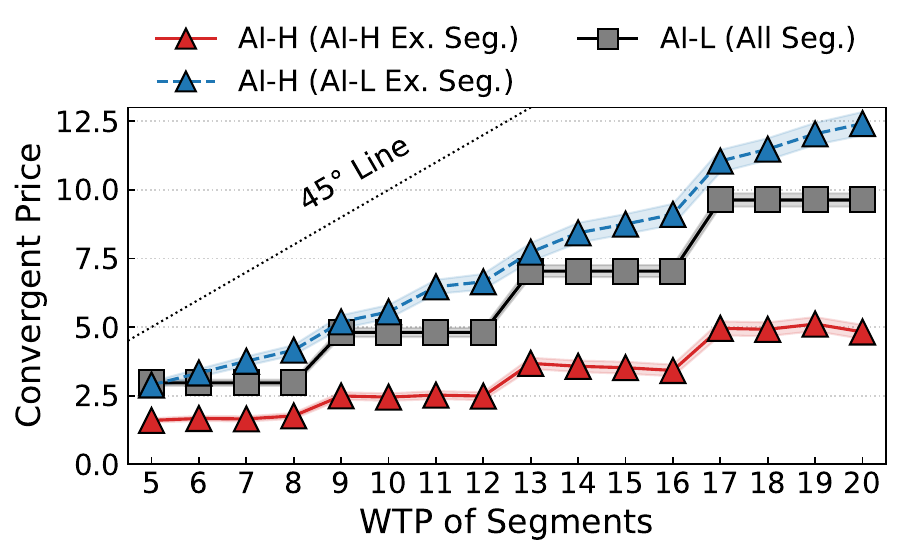}}
    \hfill
    \subfigure[$(k_H,k_L)=(16,8)$]{\includegraphics[width=0.49\textwidth]{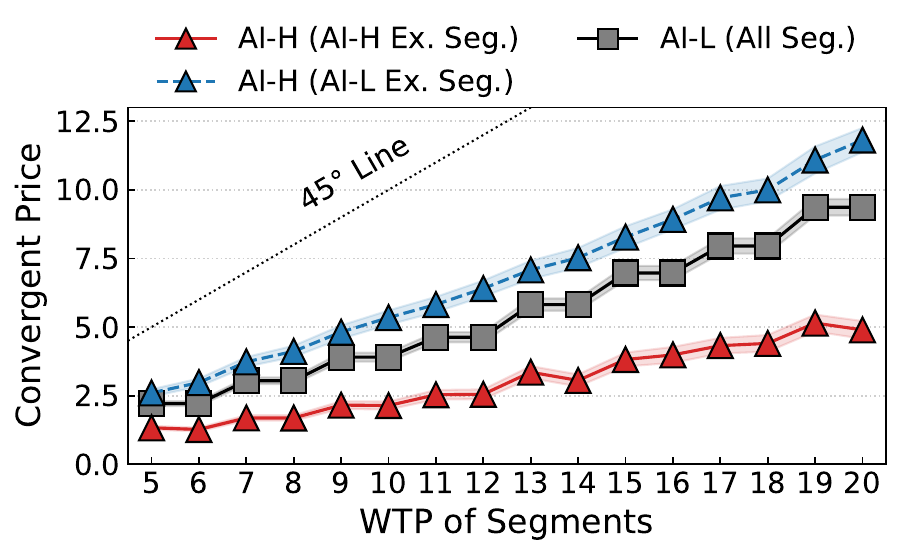}}
    
    \caption{Bait-and-Restraint-Exploit Strategy (On Average)}
    \vspace{2mm}
    \begin{minipage}{0.98\textwidth}
        \small \textit{Note:} Gray dots represent the average prices of AI-L across segments. Red dots correspond to the average prices of AI-H conditional on winning the segment, while blue dots indicate the average prices of AI-H conditional on losing the segment.\footnotemark
    \end{minipage}
    \label{fig:collusive_strategy_ave}
\end{figure}

Figure~\ref{fig:collusive_strategy_ave} reports the average prices quoted by each algorithm under the asymmetric market segmentation profiles $(16,1)$ through $(16,8)$.

First, the Bait-and-Restraint-Exploit strategy is clearly observable on average. Within the segments in AI-L's segmentation, AI-H alternates between two roles. On the one hand, it sets significantly higher ``baiting'' prices, represented by the blue dots, effectively ceding these segments to AI-L and inducing AI-L to raise its exploratory prices. On the other hand, it captures limited profits via ``exploiting'' prices, represented by the red dots, much lower than AI-L does. Notably, for the most of time, the average price quoted by AI-L lies between baiting and exploiting prices, but much closer to the former. This implies that (i) AI-H successfully baits AI-L into setting relatively high prices, and (ii) AI-H practices ``restrained exploitation'', i.e., undercutting enough to secure the segment without triggering a price collapse.

Second, we observe that the exploiting prices of AI-H are remarkably similar within segments in AI-L's segmentation. This suggests that, on average, the prices quoted by AI-H to deter downward exploration of AI-L is only conditional on AI-L's prices but independent of segments' WTPs.

\footnotetext{In panels (a) and (b), AI-L's average price appears to exceed AI-H's average baiting price, particularly in low-value segments. This pattern may seem counterintuitive because, whenever AI-H cedes a segment through baiting, its price must be strictly higher than AI-L's price in that particular instance. The apparent reversal is therefore an aggregation artifact. AI-H exploits these segments in the vast majority of observations, during which AI-L typically sets relatively high prices; these periods consequently dominate AI-L's unconditional average price. Baiting events, by contrast, are infrequent. If they occur primarily when the overall price level is relatively low, AI-H's average price conditional on baiting can fall below AI-L's unconditional average price.}

\section{Simulation Setup for the Robustness Checks in Section~6}
\subsection{Simulation Setup of Simultaneous Pricing}\label{app:simlutaneous_pricing}

Section 6.1 considers a simultaneous-pricing environment in which consumers arrive in all segments at the same time and the algorithms simultaneously set prices across segments. Buyer WTP is normalized to 1. In each segment, firms can choose among $20$ equally spaced prices: $$A := \{0.1, 0.15, \ldots, 1.0, 1.05\}.$$ A strategy is a vector $\mathbf{a} = (a_1, \ldots, a_k) \in A^k$, where $k$ is the number of segments, and each cell of the Q-vector corresponds to one strategy. In each segment, a unit of buyers purchase from the firm offering the lowest price that does not exceed their WTP; if prices are equal, the buyers split evenly between the two firms.\footnote{\label{fn:simultaneous_pricing}This assumption differs from that in the main text, where ties are broken randomly. Here, we model all buyers in each segment arriving simultaneously, reflecting a coarser time scale. In this setting, it is more reasonable to assume that a unit of buyers splits evenly between firms when prices are equal, rather than allowing a single firm to randomly capture the entire segment. Adopting the latter tie-breaking rule would generate two additional patterns.
First, the CI will increase when moving from a single segment to two segments: with a single segment, identical prices can still yield unstable outcomes due to random allocation, whereas in the two-segment case, each firm consistently captures one segment, generating positive profits and stabilizing the system. Ignoring the single-segment case, the CI still will decline as the number of segments increases.
Second, in the single-segment setting, the CI under random tie-breaking will be lower than under an even-split rule. The reason lies in stability: in a single segment, convergence requires both firms to select the same price. Under random tie-breaking, a bilateral rebound to the same high price is unstable because only one firm captures the profit. In contrast, when ties are split evenly, both firms earn positive profits, making it easier to sustain the high price after a rebound.
}
In each period $t$, firms select the strategy with the highest Q-value with probability $1-\varepsilon_t$, and choose a random strategy with probability $\varepsilon_t$.
The exploration rate $\varepsilon_t$ decays over time as in the main text. Due to computational constraints, in the simulation, we restrict attention to $k \in \{1,2,3,4\}$. All other settings are identical to the baseline specification. We conduct $100$ independent simulation runs for this specification. Table~\ref{tab:simultaneous_mkt_allocation} reports the corresponding market-allocation statistics.

\begin{table}[!h]
\centering
\begin{threeparttable}
    \caption{Statistics of Market Allocation under Simultaneous Pricing}
    \label{tab:simultaneous_mkt_allocation}
    
    \small
    \begin{tabular}{l cccc c cccc} 
    \toprule
    & \multicolumn{4}{c}{Exclusive Markets (\%)} & & \multicolumn{4}{c}{Shared Markets (\%)} \\
    \cmidrule{2-5} \cmidrule{7-10}
    Segmentation ($k$) & 1 & 2 & 3 & 4 & & 1 & 2 & 3 & 4 \\ 
    \midrule
    Total                  & 0.00 & 60.00 & 55.00 & 79.50 & & 100.00 & 40.00 & 45.00 & 20.50 \\
    Price $\ge$ 25\% WTP   & 0.00 & 60.00 & 54.33 & 79.50 & & 100.00 & 40.00 & 45.00 & 20.50 \\
    Price $\ge$ 50\% WTP   & 0.00 & 60.00 & 50.33 & 69.25 & & 100.00 & 38.50 & 44.33 & 17.75 \\
    Price $\ge$ Mean Price & 0.00 & 37.00 & 14.67 & 38.50 & & 54.00  & 24.00 & 31.33 & 13.75 \\ 
    \bottomrule
    \end{tabular}
    \begin{tablenotes}
        \small
        \item \textit{Note}: See the note to Table~1 in the main text.
    \end{tablenotes}
\end{threeparttable}
\end{table}

\subsection{Simulation Setup for Q-learning with One-Period Memory}\label{app:memory}

In Section 6.2, the state variable for algorithm $i$ at period $t$ is defined as $(p_{i,t-1},p_{-i,t-1},m_{t-1},m_{t})$, where $p_{i,t-1}$ and $p_{-i,t-1}$ denote the prices set by agent $i$ and its rival in the previous period, while $m_{t-1}$ and $m_{t}$ represent the consumer's market segments in periods $t-1$ and $t$. 
The Q-learning algorithms condition their pricing strategies on the state variable.
Consumer WTP is normalized to $1$. 
In each segment, the action space $A$ consists of $20$ equally spaced prices: $A := \{0.1, 0.15, \ldots, 1.0, 1.05\}$. 
For these simulations, we restrict our analysis to segment counts $k \in \{1,2,3,4\}$.
All other settings are identical to the baseline specification.
We conduct $100$ independent simulation runs for this specification.

\section{Additional Robustness Checks}
\subsection{Robustness-check Strategy}

\noindent\textbf{Check A: Collusion Index and the Number of Market Segments.}

Under symmetric segmentation with homogeneous consumers, we regress the collusion index (CI) on the number of market segments. Observation~2 yields the following prediction:
\vspace{-1em}
\begin{itemize}
    \item \textbf{The coefficient on the number of market segments should be negative.}
\end{itemize}

\noindent\textbf{Check B: Negative Correlation across Segments.}

Under symmetric segmentation, we compute pairwise correlations between the collusion indices (CIs) of different segments and report the proportion of segment pairs exhibiting negative correlations. Observation~2 yields the following prediction:
\vspace{-1em}
\begin{itemize}
    \item \textbf{Negative correlations should account for a majority of all pairwise correlations.}
\end{itemize}

\noindent\textbf{Check C: Collusion Index and Expected WTP.}

Under symmetric segmentation with heterogeneous consumers, we regress the segment-level collusion index (CI) on expected willingness to pay (WTP) under the \((16,16)\) market-segmentation profile. Observation~3 yields the following prediction:
\vspace{-1em}
\begin{itemize}
    \item \textbf{The coefficient on expected WTP should be positive.}
\end{itemize}

\noindent\textbf{Check D: Bait-and-Restraint-Exploit Strategy.}

Under asymmetric segmentation, we test the pricing pattern implied by Observation~5. The observation yields the following prediction:
\vspace{-1em}
\begin{itemize}
    \item Relative to AI-L, AI-H should charge a significantly \textbf{higher maximum price} but a significantly \textbf{lower minimum price}.
\end{itemize}

\subsection{Segment-Independent Action Space} \label{app:market-independent_action_space}

We alter the action space from a segment-dependent to a segment-independent setting to examine whether the results remain consistent.
Let $\overline{\omega}=20$ denote the highest WTP across all segments. 
We define the common action space $A$ as:
$$
    A := \left\{\overline{\omega}\frac{2}{l}, \overline{\omega}\frac{3}{l},\ldots, \overline{\omega}, \overline{\omega}\frac{l+1}{l}\right\},
$$
where we set $l=200$. \footnote{We apply this alternative specification to heterogeneous-consumer environments under both symmetric and asymmetric market segmentation, because with homogeneous consumers, both specifications yield the same action space across market segments and therefore the same results.}

We implement \textbf{Checks B, C under symmetric market segmentation with heterogeneous consumers(Table~\ref{tab:RC_action_space_sym}) and D under asymmetric market segmentation with heterogeneous consumers (Table~\ref{tab:RC_action_space_asym})} under this specification. 
Results show that all three predicted patterns continue to hold under the alternative specification.

\begin{table}[!h]
\centering
\begin{threeparttable}
    \caption{Segment-Independent Action Space: Symmetric Segmentation with Heterogeneous Consumers}
    \label{tab:RC_action_space_sym}
    
    \small
    \begin{tabular}{l cccc} 
    \toprule
    & \multicolumn{4}{c}{Market Segmentation Profile} \\ 
    \cmidrule{2-5}
    & (16,16) & (8,8) & (4,4) & (2,2) \\ 
    \midrule
    Negative Correlation: Proportion (\%) & 64.17 & 75  & 100 & 100 \\
    \midrule
    CI Increases w/ WTP:  Coefficient & $0.008319^*$ & & & \\
    \bottomrule
    \end{tabular}

    \begin{tablenotes}
        \small
        \item \textit{Note}: The first row reports, for each market-segmentation profile, the proportion of pairwise correlations between segment-level collusion indices that are negative. The second row reports the coefficient from regressing the segment-level collusion index on expected willingness to pay (WTP) under the $(16,16)$ segmentation profile. An asterisk ($*$) denotes statistical significance at the 5\% level ($p < 0.05$).
    \end{tablenotes}
\end{threeparttable}
\end{table}

\begin{table}[!h]
\centering
\begin{threeparttable}
    \caption{Segment-Independent Action Space: Asymmetric Segmentation with Heterogeneous Consumers}
    \label{tab:RC_action_space_asym}
    
    \small
    \begin{tabular}{l ccccc} 
    \toprule
    & \multicolumn{5}{c}{Market Segmentation Profile} \\ 
    \cmidrule{2-6}
    Statistic (t) & (16,16) & (16,8) & (16,4) & (16,2) & (16,1) \\ 
    \midrule
    $\max p_H - \max p_L$ & 0.6732  & 3.2868$^*$ & 5.5312$^*$  & 12.2912$^*$ & 17.5448$^*$ \\
    $\min p_H - \min p_L$ & --1.2492 & --6.5307$^*$ & --10.5631$^*$ & --15.3263$^*$ & --24.6788$^*$ \\
    \bottomrule
    \end{tabular}

    \begin{tablenotes}
        \small
        \item \textit{Note}: For each market segmentation profile, the first row reports the t-statistic for the difference in maximum prices between agents, while the second row reports the t-statistic for the difference in minimum prices. The tests are one-sided, with alternative hypotheses corresponding to $t > 0$ for maximum prices and $t < 0$ for minimum prices. An asterisk ($*$) indicates statistical significance at the 5\% level ($p < 0.05$).
    \end{tablenotes}
\end{threeparttable}
\end{table}

\subsection{Decay Rate}
For each market-segmentation profile, we \textbf{adjust $\beta$ to hold the expected number of explorations per cell constant at $\nu=100$}.\footnote{Given the target number of explorations $\nu$, the number of available actions $k$, and the number of states $l$, the corresponding value of $\beta$ can be determined accordingly.} 

We implement \textbf{Checks B and C under symmetric market segmentation with heterogeneous consumers (Table~\ref{tab:RC_nu_sym}) and Check D under asymmetric market segmentation with heterogeneous consumers (Table~\ref{tab:RC_nu_asym})}. Results show that the predicted patterns continue to hold when the expected number of explorations per cell is held constant.

\begin{table}[!h]
\centering
\begin{threeparttable}
    \caption{Decay Rate: Symmetric Segmentation with Heterogeneous Consumers}
    \label{tab:RC_nu_sym}
    
    \small
    \begin{tabular}{l cccc} 
    \toprule
    & \multicolumn{4}{c}{Market Segmentation Profile} \\
    \cmidrule{2-5}
    Measure & (16,16) & (8,8) & (4,4) & (2,2) \\
    \midrule
    
    Negative Correlation: Proportion (\%) & 63.33 & 71.43 & 100.00 & 100.00 \\
    
    \midrule
    
    CI increases w/ WTP: Coefficient& $0.0080^*$ & & & \\ 
    \bottomrule
    \end{tabular}

    \vspace{2mm}
    \begin{tablenotes}
       \small \item \textit{Note:} See the note to Table~\ref{tab:RC_action_space_sym}. 
    \end{tablenotes}
\end{threeparttable}
\end{table}

\begin{table}[!h]
\centering
\begin{threeparttable}
    \caption{Decay Rate: Asymmetric Segmentation with Heterogeneous Consumers}
    \label{tab:RC_nu_asym}
    
    \small
    \begin{tabular}{l ccccc} 
    \toprule
    & \multicolumn{5}{c}{Market Segmentation Profile} \\
    \cmidrule{2-6}
    Statistic ($t$) & (16,16) & (16,8) & (16,4) & (16,2) & (16,1) \\
    \midrule

    $\max p_H - \max p_L$ & --0.6441 & 0.0042 & 6.8973$^*$ & 11.2082$^*$ & 16.8340$^*$ \\
    $\min p_H - \min p_L$ & --0.1283 & --6.9163$^*$ & --9.8367$^*$ & --14.3879$^*$ & --34.2371$^*$ \\
    \bottomrule
    \end{tabular}

    \vspace{2mm}
    \begin{tablenotes}
        \small \item \textit{Note}: see the note to Table~\ref{tab:RC_action_space_asym}. Under the $(16,8)$ segmentation profile, the difference of maximum prices is not statistically significant. This exception arises because maintaining $100$ explorations per cell requires assigning AI-H a smaller value of $\beta$, which in turn prolongs its exploration phase. As a result, AI-H continues to explore with positive probability even after AI-L's pricing policy has stabilized, allowing it to undercut AI-L in some exploration rounds.
\    \end{tablenotes}
\end{threeparttable}
\end{table}

\subsection{Discount Factor}

We vary $\delta$ over the set $\{0.89, 0.91, 0.93, 0.97, 0.99\}$ to assess the robustness of our findings across alternative market-segmentation profiles.

We implement \textbf{Checks A and B under symmetric segmentation with homogeneous consumers (Table~\ref{tab:RC_delta_sym_homo}), Checks B and C under symmetric segmentation with heterogeneous consumers (Table~\ref{tab:RC_delta_sym}), and Check D under asymmetric segmentation with heterogeneous consumers(Table~\ref{tab:RC_delta_asym})}.
The results show that the predicted patterns continue to hold over this relatively high range of $\delta$.

\begin{table}[!h]
\centering
\begin{threeparttable}
    \caption{Discount Factor: Symmetric Segmentation with Homogeneous Consumers}
    \label{tab:RC_delta_sym_homo}
    
    \small

    \begin{tabular}{l cccc} 
    \toprule
    & \multicolumn{4}{c}{Market Segmentation Profile} \\
    \cmidrule{2-5}
    & (16,16) & (8,8) & (4,4) & (2,2) \\
    \midrule
    
    \multicolumn{5}{l}{\textit{Panel A: $\delta = 0.89$}} \\
    \quad Negative Correlation: Proportion (\%)             & 57.50 & 82.14 & 83.33 & 100.00 \\
    \quad  Collusion Decline: Coefficient     & \multicolumn{4}{c}{$-0.0129^*$} \\
    \addlinespace[2pt]

    \multicolumn{5}{l}{\textit{Panel B: $\delta = 0.91$}} \\
    \quad Negative Correlation: Proportion (\%)             & 61.67 & 78.57 & 83.33 & 100.00 \\
    \quad Collusion Decline: Coefficient     & \multicolumn{4}{c}{$-0.0131^*$} \\
    \addlinespace[2pt]

    \multicolumn{5}{l}{\textit{Panel C: $\delta = 0.93$}} \\
    \quad Negative Correlation: Proportion (\%)             & 57.50 & 78.57 & 100.00 & 100.00 \\
    \quad Collusion Decline: Coefficient     & \multicolumn{4}{c}{$-0.0161^*$} \\
    \addlinespace[2pt]

    \multicolumn{5}{l}{\textit{Panel D: $\delta = 0.97$}} \\
    \quad Negative Correlation: Proportion (\%)             & 68.33 & 89.29 & 100.00 & 100.00 \\
    \quad Collusion Decline: Coefficient     & \multicolumn{4}{c}{$-0.0244^*$} \\
    \addlinespace[2pt]

    \multicolumn{5}{l}{\textit{Panel E: $\delta = 0.99$}} \\
    \quad Negative Correlation: Proportion (\%)             & 68.33 & 89.29 & 100.00 & 100.00 \\
    \quad Collusion Decline: Coefficient     & \multicolumn{4}{c}{$-0.0282^*$} \\
    \bottomrule
    \end{tabular}

    \begin{tablenotes}
        \small
        \item \textit{Note}: For each $\delta$ and each market segmentation profile, the first row shows the proportion of negative correlations among all correlation coefficients. The second row reports the regression coefficient of CIs on the number of segments. An asterisk ($*$) denotes statistical significance at the 5\% level ($p < 0.05$). 
    \end{tablenotes}
\end{threeparttable}
\end{table}

\begin{table}[!h]
\centering
\begin{threeparttable}
    \caption{Discount Factor: Symmetric Segmentation with Heterogeneous Consumers}
    \label{tab:RC_delta_sym}
    
    \small

    \begin{tabular}{l cccc} 
    \toprule
    & \multicolumn{4}{c}{Market Segmentation Profile} \\
    \cmidrule{2-5}
    & (16,16) & (8,8) & (4,4) & (2,2) \\ 
    \midrule
    
    \multicolumn{5}{l}{\textit{Panel A: $\delta = 0.89$}} \\
    \quad Negative Correlation: Proportion (\%)             & 58.33 & 60.71 & 100.00 & 100.00 \\
    \quad CI increases w/ WTP: Coefficient & $0.0047^*$ & & & \\
    \addlinespace[2pt]

    \multicolumn{5}{l}{\textit{Panel B: $\delta = 0.91$}} \\
    \quad Negative Correlation: Proportion (\%)             & 65.00 & 75.00 & 83.33 & 100.00 \\
    \quad CI increases w/ WTP: Coefficient & $0.0044^*$ & & & \\
    \addlinespace[2pt]

    \multicolumn{5}{l}{\textit{Panel C: $\delta = 0.93$}} \\
    \quad Negative Correlation: Proportion (\%)             & 59.17 & 60.21 & 66.67 & 100.00 \\
    \quad CI increases w/ WTP: Coefficient & $0.0054^*$ & & & \\
    \addlinespace[2pt]

    \multicolumn{5}{l}{\textit{Panel D: $\delta = 0.95$}} \\
    \quad Negative Correlation: Proportion (\%)             & 68.33 & 75.00 & 66.37 & 100.00 \\
    \quad CI increases w/ WTP: Coefficient & $0.0062^*$ & & & \\
    \addlinespace[2pt]

    \multicolumn{5}{l}{\textit{Panel E: $\delta = 0.97$}} \\
    \quad Negative Correlation: Proportion (\%)             & 63.33 & 82.14 & 100.00 & 100.00 \\
    \quad CI increases w/ WTP: Coefficient & $0.0071^*$ & & & \\
    \addlinespace[2pt]

    \multicolumn{5}{l}{\textit{Panel F: $\delta = 0.99$}} \\
    \quad Negative Correlation: Proportion (\%)             & 65.83 & 71.43 & 83.33 & 100.00 \\
    \quad CI increases w/ WTP: Coefficient & $0.0107^*$ & & & \\
    \bottomrule
    \end{tabular}

    \begin{tablenotes}
        \small
        \item \textit{Note}: see the note to Table~\ref{tab:RC_action_space_sym}. 
    \end{tablenotes}
\end{threeparttable}
\end{table}

\begin{table}[!h]
\centering
\begin{threeparttable}
    \caption{Discount Factor: Asymmetric Segmentation with Heterogeneous Consumers}
    \label{tab:RC_delta_asym}
    
    \small

    \begin{tabular}{l ccccc} 
    \toprule
    & \multicolumn{5}{c}{Market Segmentation Profile} \\
    \cmidrule{2-6}
    Statistic ($t$) & (16,16) & (16,8) & (16,4) & (16,2) & (16,1) \\
    \midrule
    
    \multicolumn{6}{l}{\textit{Panel A: $\delta = 0.89$}} \\
    \quad $\max p_H - \max p_L$ & --0.4062 & 2.6984$^*$ & 6.4050$^*$ & 13.5245$^*$ & 19.3174$^*$ \\
    \quad $\min p_H - \min p_L$ & 0.4033 & --7.6181$^*$ & --10.2003$^*$ & --16.7344$^*$ & --21.9277$^*$ \\
    \addlinespace[2pt]

    \multicolumn{6}{l}{\textit{Panel B: $\delta = 0.91$}} \\
    \quad $\max p_H - \max p_L$ & --1.6238 & 4.0637$^*$ & 8.1924$^*$ & 10.6634$^*$ & 20.4144$^*$ \\
    \quad $\min p_H - \min p_L$ & 0.2590 & --5.7212$^*$ & --11.3310$^*$ & --19.3366$^*$ & --29.3229$^*$ \\
    \addlinespace[2pt]

    \multicolumn{6}{l}{\textit{Panel C: $\delta = 0.93$}} \\
    \quad $\max p_H - \max p_L$ & --0.9335 & 3.9488$^*$ & 6.3088$^*$ & 10.6573$^*$ & 18.1625$^*$ \\
    \quad $\min p_H - \min p_L$ & --0.0053 & --7.4492$^*$ & --9.9648$^*$ & --14.9194$^*$ & --28.4024$^*$ \\
    \addlinespace[2pt]

    \multicolumn{6}{l}{\textit{Panel D: $\delta = 0.95$}} \\
    \quad $\max p_H - \max p_L$ & --1.0887 & 3.0278$^*$ & 7.7210$^*$ & 11.7114$^*$ & 19.7238$^*$ \\
    \quad $\min p_H - \min p_L$ & --0.4525 & --6.7320$^*$ & --10.9716$^*$ & --16.6637$^*$ & --32.6932$^*$ \\
    \addlinespace[2pt]

    \multicolumn{6}{l}{\textit{Panel E: $\delta = 0.97$}} \\
    \quad $\max p_H - \max p_L$ & 0.7361 & 4.0600$^*$ & 7.8321$^*$ & 12.0779$^*$ & 17.1495$^*$ \\
    \quad $\min p_H - \min p_L$ & --0.6419 & --5.6294$^*$ & --10.4340$^*$ & --16.2057$^*$ & --33.8795$^*$ \\
    \addlinespace[2pt]

    \multicolumn{6}{l}{\textit{Panel F: $\delta = 0.99$}} \\
    \quad $\max p_H - \max p_L$ & 0.0899 & 3.5235$^*$ & 6.0150$^*$ & 11.7388$^*$ & 16.7701$^*$ \\
    \quad $\min p_H - \min p_L$ & --0.4581 & --4.8961$^*$ & --9.9379$^*$ & --15.7006$^*$ & --32.3249$^*$ \\
    \bottomrule
    \end{tabular}

    \vspace{2mm}
    \begin{tablenotes}
        \small \item \textit{Note}: see the note to Table~\ref{tab:RC_action_space_asym}.
    \end{tablenotes}
\end{threeparttable}
\end{table}

\subsection{Learning Rate}

We vary $\alpha$ over the set $\{0.05, 0.1, 0.2\}$ to assess the robustness of our findings across alternative market-segmentation profiles.

We implement \textbf{Checks A and B under symmetric segmentation with homogeneous consumers (Table~\ref{tab:RC_alpha_sym_homo}), Checks B and C under symmetric segmentation with heterogeneous consumers (Table~\ref{tab:RC_alpha_sym}), and Check D under asymmetric segmentation with heterogeneous consumers(Table~\ref{tab:RC_alpha_asym})}.
The results show that the predicted patterns continue to hold over this relatively low range of $\alpha$.

\begin{table}[!h]
\centering
\begin{threeparttable}
    \caption{Learning Rate: Symmetric Segmentation with Homogeneous Consumers}
    \label{tab:RC_alpha_sym_homo}
    
    \small
    \begin{tabular}{l cccc} 
    \toprule
    & \multicolumn{4}{c}{Market Segmentation Profile} \\
    \cmidrule{2-5}
    & (16,16) & (8,8) & (4,4) & (2,2) \\ 
    \midrule
    
    \multicolumn{5}{l}{\textit{Panel A: $\alpha = 0.05$}} \\
    \quad Negative Correlation: Proportion (\%)             & 68.33 & 75.00 & 100.00 & 100.00 \\
    \quad Collusion Decline: Coefficient     & \multicolumn{4}{c}{$-0.0159^*$} \\
    \addlinespace[2pt]

    \multicolumn{5}{l}{\textit{Panel B: $\alpha = 0.1$}} \\
    \quad Negative Correlation: Proportion (\%)             & 71.67 & 89.29 & 83.33 & 100.00 \\
    \quad Collusion Decline: Coefficient     & \multicolumn{4}{c}{$-0.0218^*$} \\
    \addlinespace[2pt]

    \multicolumn{5}{l}{\textit{Panel C: $\alpha = 0.2$}} \\
    \quad Negative Correlation: Proportion (\%)             & 62.50 & 67.86 & 83.33 & 100.00 \\
    \quad Collusion Decline: Coefficient     & \multicolumn{4}{c}{$-0.0169^*$} \\
    \bottomrule
    \end{tabular}

    \begin{tablenotes}
        \small
        \item \textit{Note}: see the note to Table~\ref{tab:RC_delta_sym_homo}.
    \end{tablenotes}
\end{threeparttable}
\end{table}

\begin{table}[!h]
\centering
\begin{threeparttable}
    \caption{Learning Rate: Symmetric Segmentation with Heterogeneous Consumers}
    \label{tab:RC_alpha_sym}
    
    \small

    \begin{tabular}{l cccc} 
    \toprule
    & \multicolumn{4}{c}{Market Segmentation Profile} \\
    \cmidrule{2-5}
    & (16,16) & (8,8) & (4,4) & (2,2) \\
    \midrule
    
    \multicolumn{5}{l}{\textit{Panel A: $\alpha = 0.05$}} \\
    \quad Negative Correlation: Proportion (\%)             & 72.50 & 82.14 & 83.33 & 100.00 \\
    \quad CI increases w/ WTP: Coefficient & $0.0153^*$ & & & \\
    \addlinespace[2pt]

    \multicolumn{5}{l}{\textit{Panel B: $\alpha = 0.1$}} \\
    \quad Negative Correlation: Proportion (\%)             & 65.83 & 67.86 & 100.00 & 100.00 \\
    \quad CI increases w/ WTP: Coefficient & $0.0107^*$ & & & \\
    \addlinespace[2pt]

    \multicolumn{5}{l}{\textit{Panel C: $\alpha = 0.15$}} \\
    \quad Negative Correlation: Proportion (\%)             & 60.83 & 67.86 & 83.33 & 100.00 \\
    \quad CI increases w/ WTP: Coefficient & $0.0077^*$ & & & \\
    \addlinespace[2pt]

    \multicolumn{5}{l}{\textit{Panel D: $\alpha = 0.2$}} \\
    \quad Negative Correlation: Proportion (\%)             & 63.33 & 75.00 & 100.00 & 100.00 \\
    \quad CI increases w/ WTP: Coefficient & $0.0069^*$ & & & \\
    \bottomrule
    \end{tabular}

    \vspace{2mm}
    \begin{tablenotes}
        \small
        \item \textit{Note}: see the note to Table~\ref{tab:RC_action_space_sym}.
     \end{tablenotes}
\end{threeparttable}
\end{table}

\begin{table}[!h]
\centering
\begin{threeparttable}
    \caption{Learning Rate: Asymmetric Segmentation with Heterogeneous Consumers}
    \label{tab:RC_alpha_asym}
    
    \small

    \begin{tabular}{l ccccc} 
    \toprule
    & \multicolumn{5}{c}{Market Segmentation Profile} \\
    \cmidrule{2-6}
    Statistic ($t$) & (16,16) & (16,8) & (16,4) & (16,2) & (16,1) \\
    \midrule
    
    \multicolumn{6}{l}{\textit{Panel A: $\alpha = 0.05$}} \\
    \quad $\max p_H - \max p_L$ & 0.2582 & 2.2909$^*$ & 5.1340$^*$ & 8.8902$^*$ & 14.8711$^*$ \\
    \quad $\min p_H - \min p_L$ & --0.5399 & --5.9413$^*$ & --7.1434$^*$ & --10.7362$^*$ & --29.8193$^*$ \\
    \addlinespace[2pt]

    \multicolumn{6}{l}{\textit{Panel B: $\alpha = 0.1$}} \\
    \quad $\max p_H - \max p_L$ & --0.5252 & 3.7393$^*$ & 6.6179$^*$ & 13.3371$^*$ & 17.6680$^*$ \\
    \quad $\min p_H - \min p_L$ & 0.2865 & --3.9846$^*$ & --9.8691$^*$ & --14.3853$^*$ & --33.9572$^*$ \\
    \addlinespace[2pt]

    \multicolumn{6}{l}{\textit{Panel C: $\alpha = 0.15$}} \\
    \quad $\max p_H - \max p_L$ & --0.7308 & 2.2345$^*$ & 6.6221$^*$ & 13.3781$^*$ & 19.4092$^*$ \\
    \quad $\min p_H - \min p_L$ & 0.7158 & --6.6827$^*$ & --10.8737$^*$ & --16.0078$^*$ & --34.7897$^*$ \\
    \addlinespace[2pt]

    \multicolumn{6}{l}{\textit{Panel D: $\alpha = 0.2$}} \\
    \quad $\max p_H - \max p_L$ & --1.1507 & 1.6785$^*$ & 8.2430$^*$ & 13.4191$^*$ & 19.6346$^*$ \\
    \quad $\min p_H - \min p_L$ & 1.7108 & --6.5102$^*$ & --10.7058$^*$ & --19.8970$^*$ & --29.8310$^*$ \\
    \bottomrule
    \end{tabular}

    \vspace{2mm}
    \begin{tablenotes}
        \small \item \textit{Note}: see the note to Table~\ref{tab:RC_action_space_asym}.
    \end{tablenotes}
\end{threeparttable}
\end{table}

\section{Declining and Uneven Collusion across Heterogeneous Segments}

\begin{figure}[!h]
    \centering
    \includegraphics[width=0.5\textwidth]{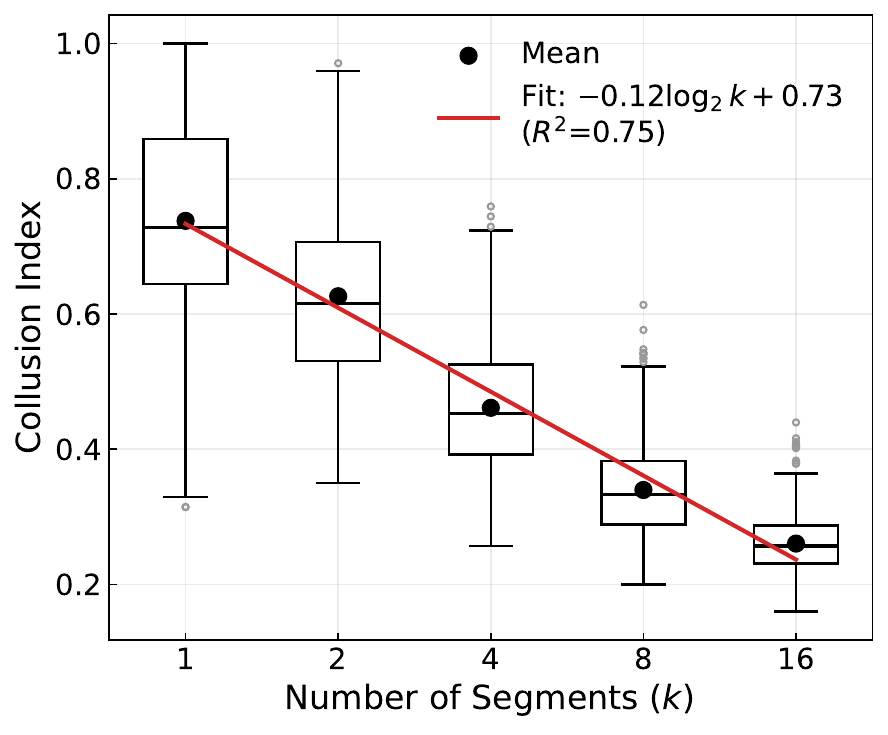}
    \caption{Declining Collusion under Heterogeneous Segments}
    \label{fig:collusion_incline_heterogeneous_segments}
\end{figure}

\begin{figure}[!h]
    \centering
    \subfigure[Segmentation Profile (16,16)]{\includegraphics[width=0.3\textwidth]{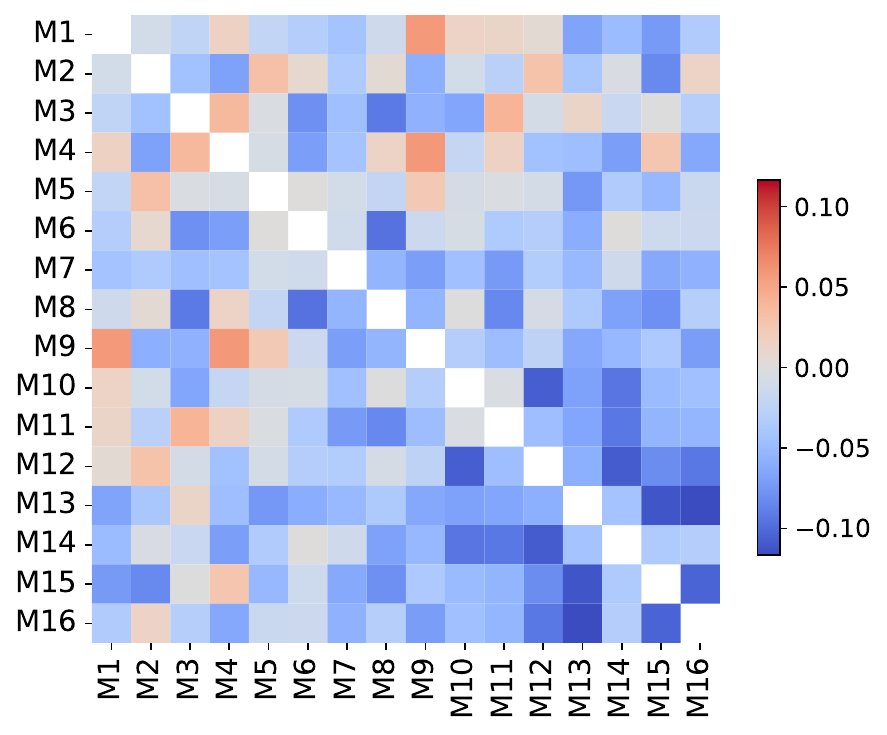}}
    \subfigure[Segmentation Profile (8,8)]{\includegraphics[width=0.3\textwidth]{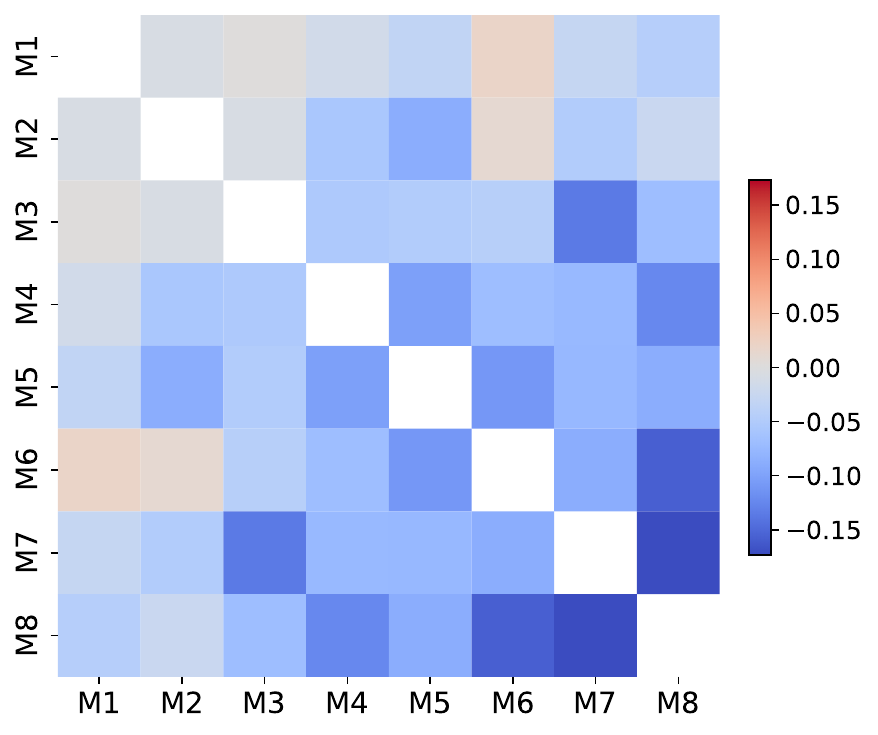}}\\
    \subfigure[Segmentation Profile (4,4)]{\includegraphics[width=0.3\textwidth]{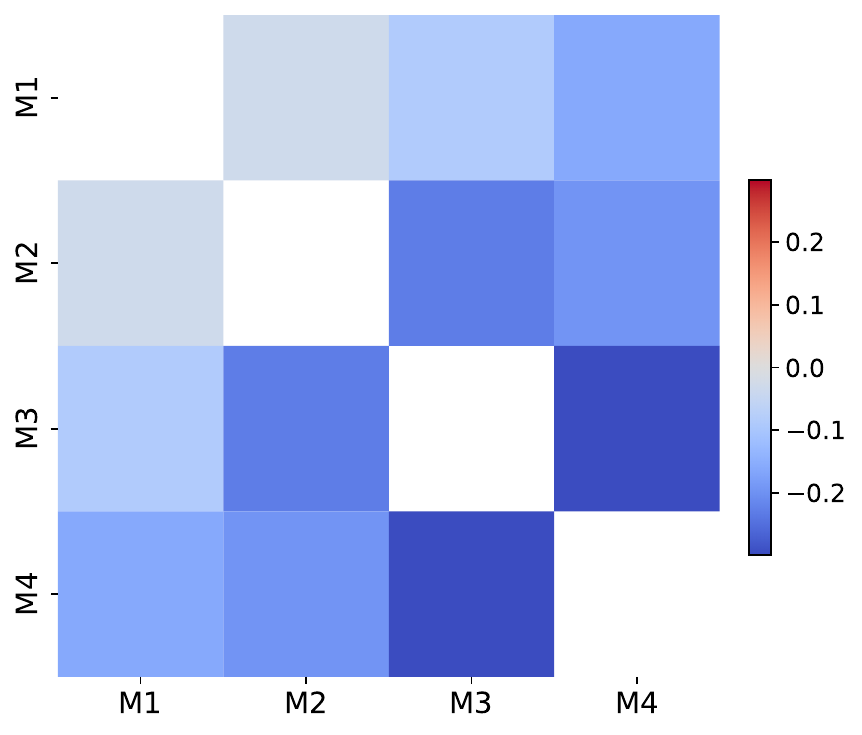}}
    \subfigure[Segmentation Profile (2,2)]{\includegraphics[width=0.3\textwidth]{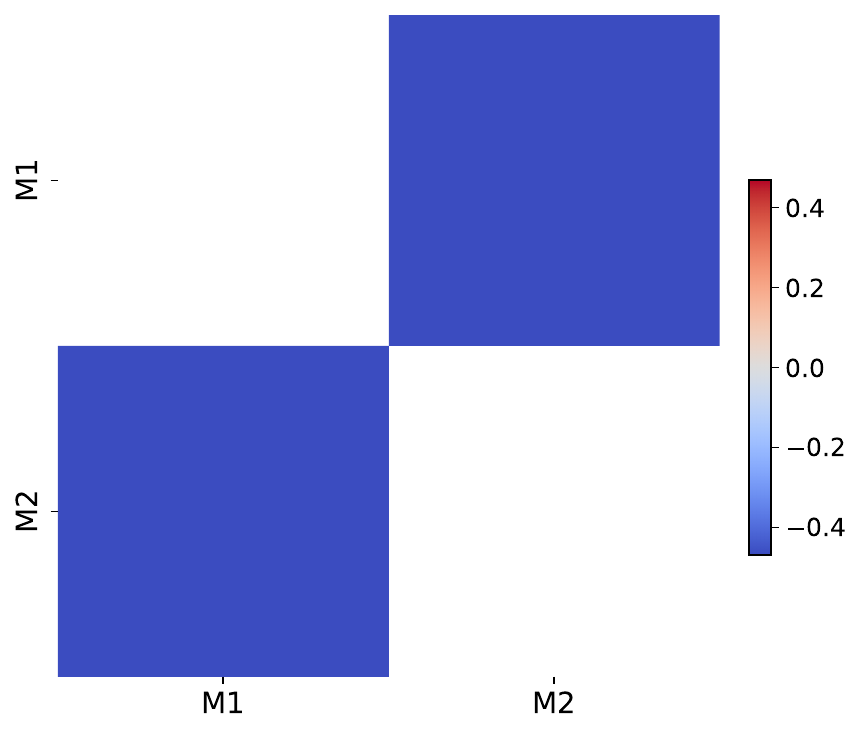}}
    \caption{Uneven Collusion under Heterogeneous Segments}
    \label{fig:negative_collusion_heterogeneous_segments}
\end{figure}

Figure~\ref{fig:collusion_incline_heterogeneous_segments} and \ref{fig:negative_collusion_heterogeneous_segments} show that the main result established for homogeneous segments, summarized in Observation~2, continues to hold when segments are heterogeneous.

\section{State–Action–Reward–State–Action}

\subsection{Setup}
As a robustness check, we replace Q-learning with the State--Action--Reward--State--Action (SARSA) algorithm while keeping all other elements of the simulation unchanged. The key distinction is that Q-learning is an off-policy algorithm, whereas SARSA is an on-policy algorithm. In the baseline Q-learning specification, the continuation value is determined by the maximum Q-value available in the next-period segment. Under SARSA, by contrast, the continuation value is determined by the Q-value of the action actually selected in the next-period segment under the prevailing exploration policy.

Specifically, firm i's Q-value is updated according to
$$
\begin{aligned}
Q_{i,t+1}(m_{i,t},p_{i,t})
={}&
(1-\alpha)Q_{i,t}(m_{i,t},p_{i,t}) \
&+\alpha
\left[
r_{i,t}
+\delta Q_{i,t}(m_{i,t+1},p_{i,t+1})
\right],
\end{aligned}
$$
where $m_{i,t}$ and $p_{i,t}$ denote firm (i)'s realized segment and selected price in period (t), respectively; $r_{i,t}$ is its realized current-period profit; $m_{i,t+1}$ is the segment realized in period (t+1); and $p_{i,t+1}$ is the price actually selected in that segment according to the firm's prevailing exploration policy.

Accordingly, SARSA replaces the Q-learning continuation term
$$
\max_{p\in A_i(m_{i,t+1})}
Q_{i,t}(m_{i,t+1},p)
$$
with
$$
Q_{i,t}(m_{i,t+1},p_{i,t+1}).
$$
The SARSA update therefore reflects the firm's realized state--action transition, including exploratory actions, rather than evaluating the next state under the currently optimal action.

Although SARSA uses a different updating rule, it retains two features of Q-learning that are central to our analysis. 
First, as the exploration rate becomes sufficiently small, SARSA selects the action with the highest Q-value in the next-period segment with high probability. Its continuation term therefore approaches the maximum next-state Q-value used by Q-learning. This similarity is particularly important for sustaining collusion in the single-segment environment. 
Second, in a multi-segment environment, the Q-value of the action selected in the next-period segment enters the update of the current segment's Q-value. SARSA therefore also generates cross-segment Q-value spillovers, which are the primary mechanism underlying our main results.

We adopt the same setup as in Section~4.1, which considers heterogeneous segments under symmetric market segmentation. 

\subsection{Results}

We assess the robustness of Observations~1--3 in the main text.

Table \ref{tab:sarsa} shows that Observation~1 remain robust.

\begin{table}[!h]
\centering
\begin{threeparttable}
    \caption{Statistics of Market Allocation under SARSA}
    \label{tab:sarsa}
    
    \small
    \begin{tabular}{l ccccc c ccccc} 
    \toprule
    & \multicolumn{5}{c}{Exclusive Markets (\%)} & & \multicolumn{5}{c}{Shared Markets (\%)} \\
    \cmidrule{2-6} \cmidrule{8-12}
    Segmentation ($k$) & 1 & 2 & 4 & 8 & 16 & & 1 & 2 & 4 & 8 & 16 \\ 
    \midrule
    Total                  & 0.00 & 94.50 & 90.50 & 88.88 & 88.19 & & 100.00 & 5.50 & 9.50 & 11.12 & 11.81 \\
    Price $\ge$ 25\% WTP   & 0.00 & 84.50 & 59.25 & 42.75 & 28.87 & & 100.00 & 5.00 & 9.25 & 10.25 & 9.63 \\
    Price $\ge$ 50\% WTP   & 0.00 & 41.00 & 26.25 & 17.00 & 12.19 & & 32.00 & 3.50 & 5.75 & 5.38 & 5.56 \\
    Price $\ge$ Mean Price & 0.00 & 37.50 & 38.50 & 34.25 & 28.63 & & 38.00  & 4.50 & 8.50 & 9.13 & 9.13 \\ 
    \bottomrule
    \end{tabular}
    \begin{tablenotes}
        \small
        \item \textit{Note}: See the note to Table~1 in the main text.
    \end{tablenotes}
\end{threeparttable}
\end{table}

\begin{figure}[!h]
    \centering
    \includegraphics[width=0.5\textwidth]{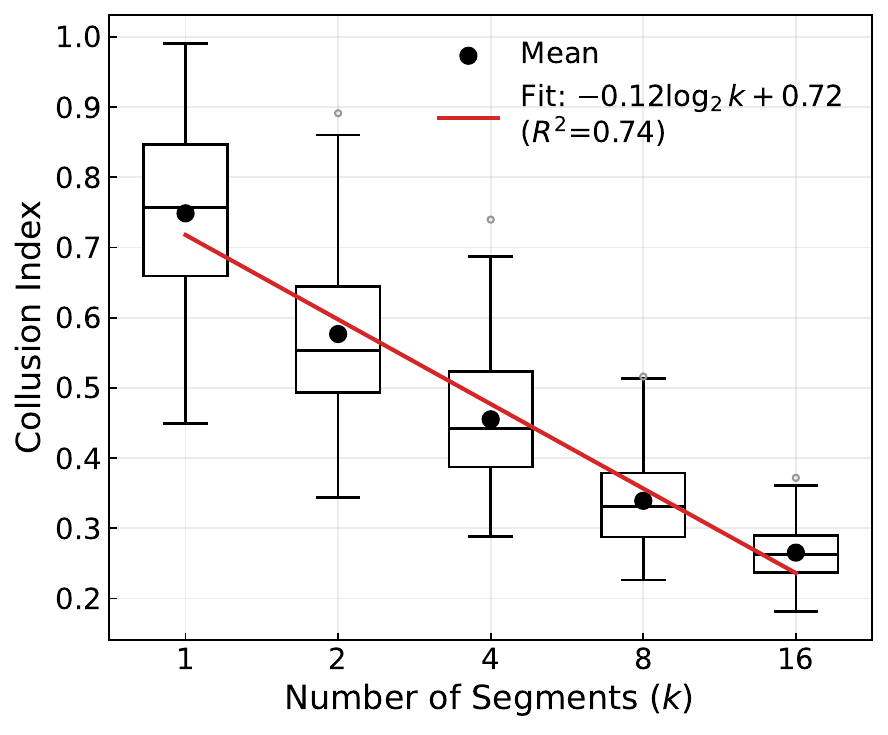}
    \caption{Declining Collusion under SARSA}
    \label{fig:collusion_incline_sarsa}
\end{figure}

\begin{figure}[!h]
    \centering
    \subfigure[Segmentation Profile (16,16)]{\includegraphics[width=0.3\textwidth]{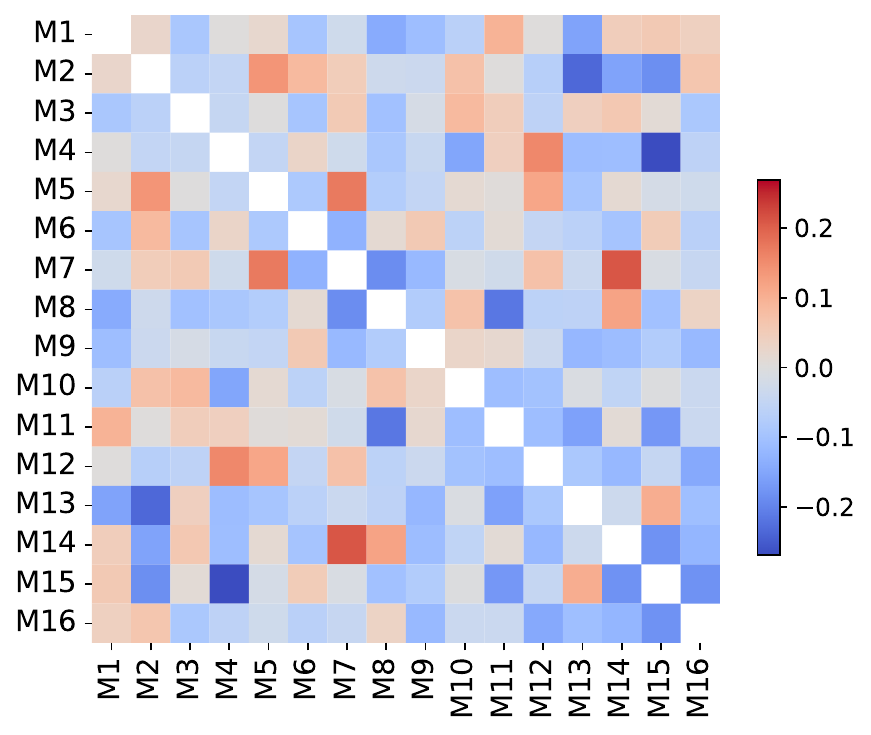}}
    \subfigure[Segmentation Profile (8,8)]{\includegraphics[width=0.3\textwidth]{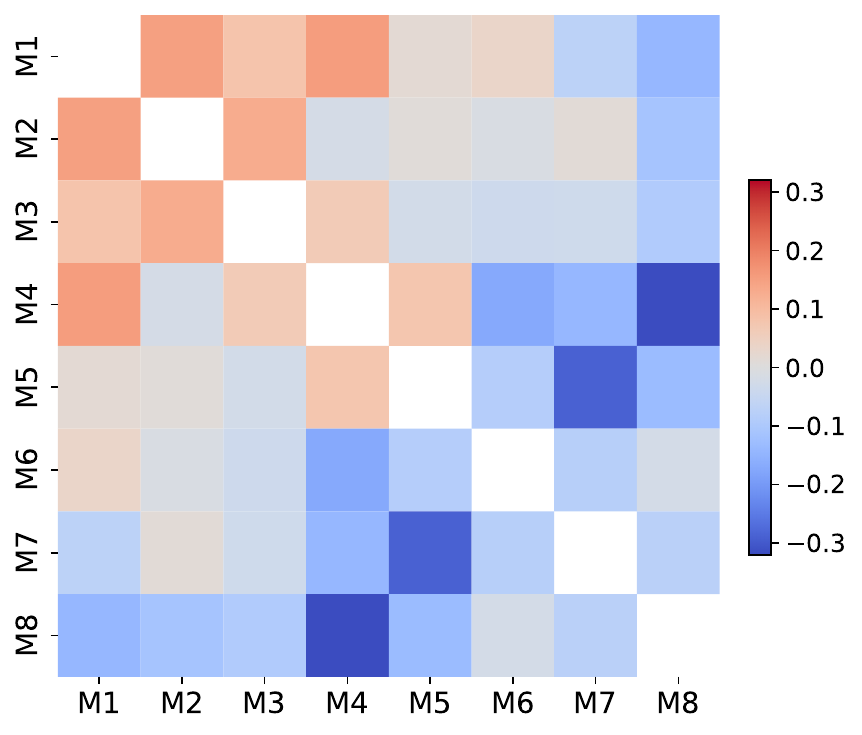}}\\
    \subfigure[Segmentation Profile (4,4)]{\includegraphics[width=0.3\textwidth]{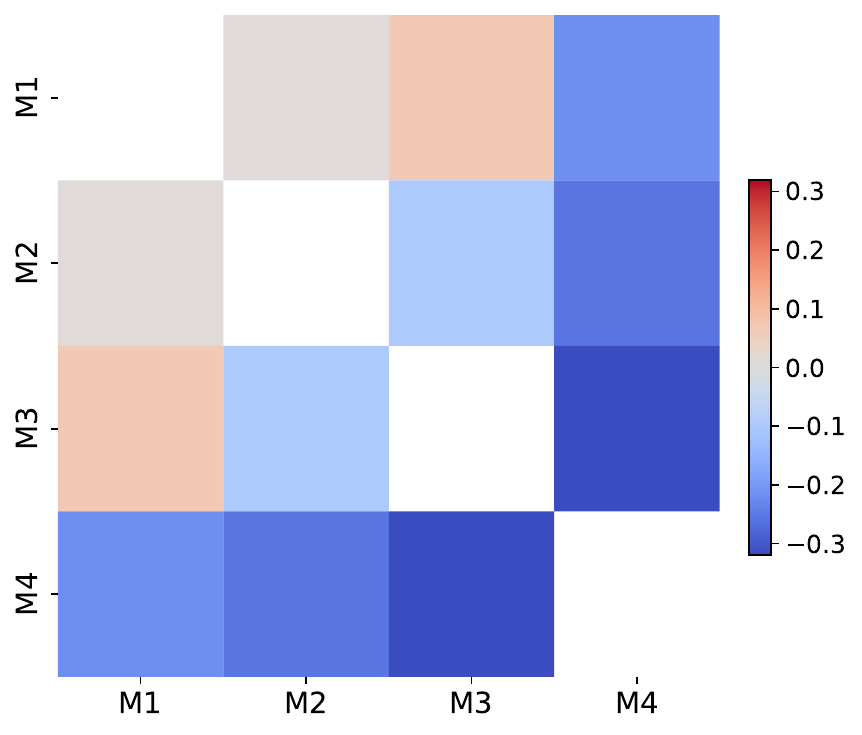}}
    \subfigure[Segmentation Profile (2,2)]{\includegraphics[width=0.3\textwidth]{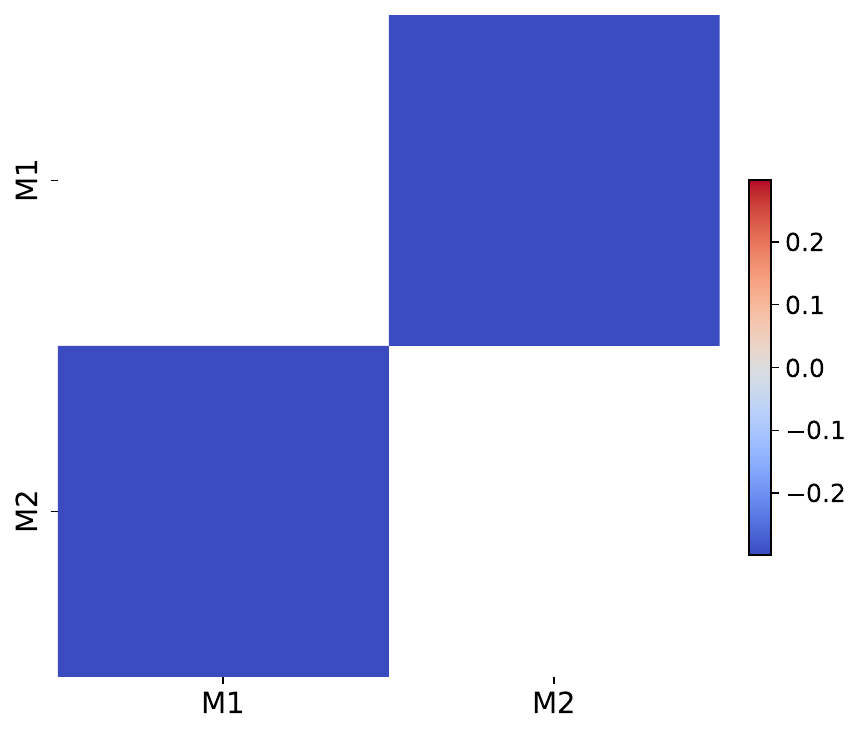}}
    \caption{Uneven Collusion under SARSA}
    \label{fig:negative_collusion_sarsa}
\end{figure}

Figures~\ref{fig:collusion_incline_sarsa} and~\ref{fig:negative_collusion_sarsa} show that Observation~2 remains robust under SARSA. Although Figure~\ref{fig:negative_collusion_sarsa} contains some positive correlations, these do not contradict our analysis. Most of these positive correlations occur between low-value segments and arise when both segments exhibit similarly intense competition, consistent with our analysis in Section~3.2.

\begin{figure}[!h]
    \centering
    \includegraphics[width=0.7\linewidth]{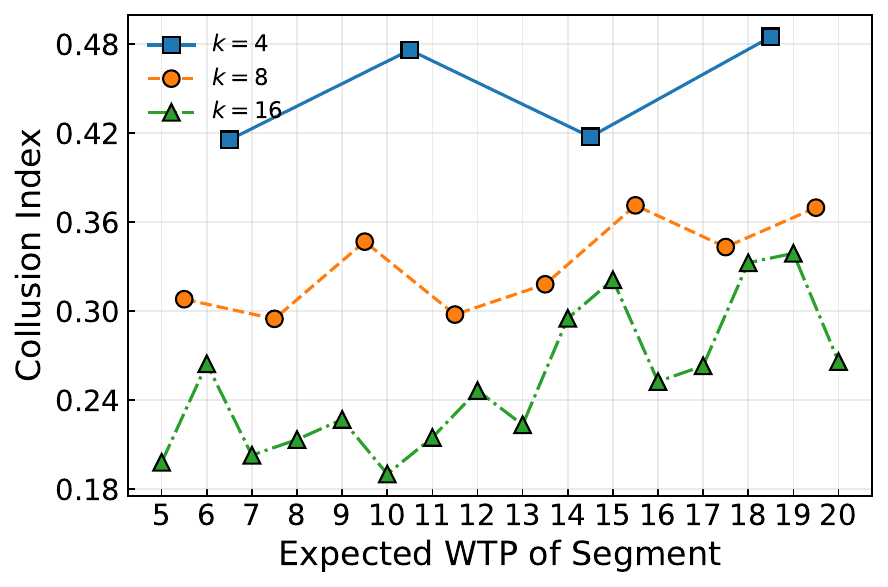}
    \caption{Variation in CI across Segments under SARSA}
    \label{fig:CI_WTP_sarsa}
\end{figure}

Figure~\ref{fig:CI_WTP_sarsa} shows that Observation~3 remains robust under SARSA.

